%% file: main.tex
\documentclass[10pt,twocolumn,letterpaper]{article}

\usepackage{cvpr}              

\input{preamble}

\definecolor{cvprblue}{rgb}{0.21,0.49,0.74}
\usepackage[pagebackref,breaklinks,colorlinks,allcolors=cvprblue]{hyperref}
\usepackage[accsupp]{axessibility}

\def\paperID{1234} 
\def\confName{CVPR}
\def\confYear{2026}

\title{ARS-Avatar: Animatable and Relightable Surfel Avatars with Learnable Ambient Occlusion}

\author{
Jiateng Liu$^1$, 
Hao Gao$^1$$^*$, 
Junxin Sun$^1$, 
Mengqi Liu$^1$, 
Jiu-Cheng Xie$^1$, 
Jucheng Song$^3$,
Feng Xu$^2$$^*$\\
$^1$Nanjing University of Posts and Telecommunications \\
$^2$Tsinghua University \quad
$^3$Macao Polytechnic University \\
\url{https://arsavatar.github.io/}
}

\begin{document}

\input{figure-src/teaser}

\let\thefootnote\relax\footnote{$*$Corresponding author: Hao Gao and Feng Xu.}

\input{sec/0_abstract}    
\input{sec/1_intro}
\input{sec/2_relate}
\input{sec/3_method}
\input{sec/4_experiment}
\input{sec/5_discuss}
{
    \small
    \bibliographystyle{ieeenat_fullname}
    \bibliography{main}
}

\clearpage
{\Large \noindent\textbf{Supplemental Document}\\}
\input{sec/6_supp}
\end{document}

%% file: preamble.tex
\usepackage{graphicx}
\usepackage{float}
\usepackage{amsmath}
\usepackage{amssymb}
\usepackage{multirow}
\usepackage{booktabs}



%% file: figure-src/teaser.tex
\twocolumn[{
\maketitle
\begin{figure}[H]
\hsize=\textwidth
\centering
\includegraphics[width=1.7\linewidth]{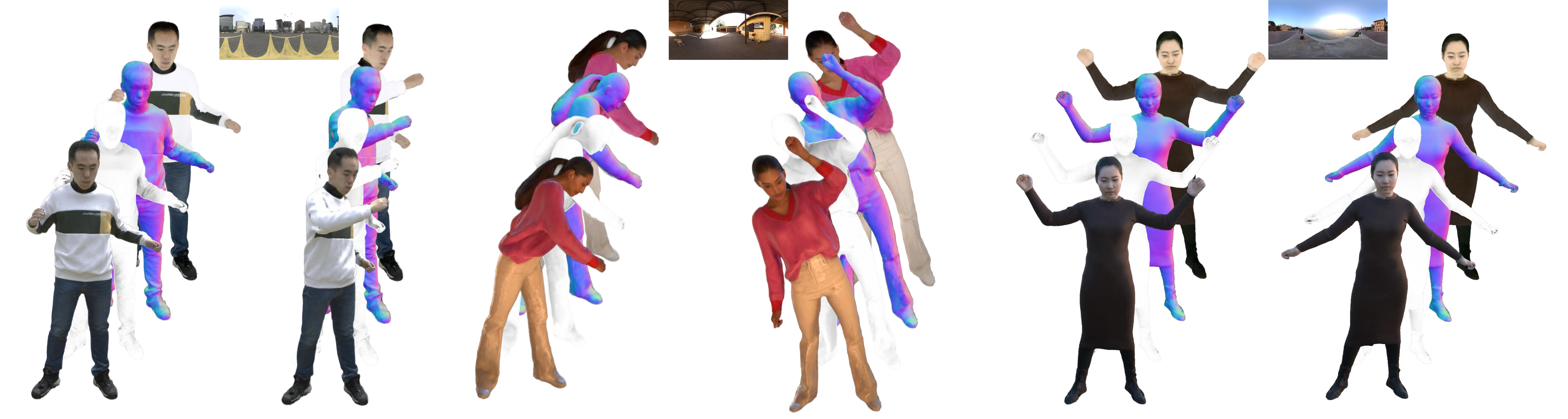}
\caption{\textbf{ARS-Avatar} creates animatable and relightable human avatars with surfels. Given novel poses and illuminations, it reconstructs high-fidelity \textit{appearance}, \textit{geometry}, \textit{BRDF materials}, and \textbf{\textit{optimizable occlusion}} for realistic animation and relighting.}
\label{fig:teaser}
\end{figure}
}]

%% file: sec/0_abstract.tex
\begin{abstract} 
Creating animatable and relightable human avatars from multi-view images remains challenging, as pose-dependent deformation, materials, and light visibility are tightly coupled in images.
In this paper, we present \textbf{ARS-Avatar}, a novel method using surfel representation for high-quality, animatable, and relightable human avatars from multi-view images captured under unknown illumination.
For faithful surfel estimation, we extract deformation displacement and body model deviation two priors from the mesh, which jointly capture local deformation complexity and serve as auxiliary.
We employ deferred shading to estimate BRDF materials and further introduce a differentiable screen-space ambient occlusion formulation that enables gradient-based optimization of body-part specific occlusion radii through finite differences, providing an efficient approximation of light visibility that can be jointly optimized with the avatar.
Extensive experiments show that ARS-Avatar achieves competitive or improved radiance reconstruction on multiple metrics and consistently outperforms the evaluated PBR relighting baselines,
while enabling realistic animation and relighting under novel poses and illuminations.
\end{abstract}

%% file: sec/1_intro.tex
\section{Introduction}
\label{sec:intro}
Creating animatable and relightable human avatars remains a fundamental problem in computer graphics, with applications in embodied AI (EAI), VR/AR, virtual try-on, and telepresence. Traditional methods rely on high-quality 3D scans and calibrated illumination~\cite{guo2019relightables, debevec2000acquiring}. However, modeling such avatars solely from multi-view images captured under unknown illumination remains ill-posed, as geometry, materials, and illumination are tightly coupled.

Recent advances in neural rendering~\cite{mildenhall2021nerf, kerbl20233d} have substantially improved high-fidelity human avatar reconstruction from multi-view images. 
Although NeRF-based methods~\cite{xiao2024neca,wu2025fast,lin2024relightable,xu2024relightable} achieve impressive quality, expensive volume rendering limits their practicality, which motivates efficient explicit primitives~\cite{jiang2025dnf}. 
Building on 3DGS, recent methods~\cite{zhao2025surfel, jiang2025dnf, li2023animatable, wang2025relightable, zhan2025interactive, choi2025relightable} reconstruct animatable and relightable avatars by combining primitives with templates~\cite{loper2015smpl, pavlakos2019smplx} and physically-based rendering~\cite{burley2012physically, rendering2015physically}.
However, capturing the rich pose-dependent deformation of clothed humans remains challenging. The local deformation complexity is highly subject-specific and depends on the interplay between body, cloth, and pose. Although existing methods introduce auxiliary features to improve pose-dependent deformation, they rarely consider the complexity and subject-specific nature of local deformation.
Furthermore, although visibility estimation has been explored for meshes~\cite{zhan2025interactive, chen2024meshavatar} and volume rendering~\cite{wang2024intrinsic, xiao2024neca}, extending it to discrete Gaussian surfels is non-trivial.
These limitations jointly constrain the achievable fidelity of pose-dependent animation and relighting.

Therefore, we present \textbf{ARS-Avatar}, an animatable and relightable human avatar represented by GaussianSurfels~\cite{dai2024high} that achieves high-fidelity rendering under novel poses and illuminations.
Specifically, we employ SMPL-X~\cite{pavlakos2019smplx} to construct the canonical avatar and reconstruct appearance, geometry, and BRDF materials under a driving pose.
The driving pose is encoded into a position map~\cite{li2024animatable} and fed into StyleUNets to estimate surfel attributes.
We further extract two deformation priors from the template mesh, i.e., the deformation displacement and the deviation from the parametric body model, both of which reflect local deformation complexity. 
The extracted priors are incorporated into surfel attribute estimation to produce more faithful pose-dependent details.
Furthermore,
we employ screen-space ambient occlusion (SSAO) as an efficient approximation and incorporate it into the optimization via a finite-difference formulation with respect to the occlusion radius. Each human body part has an initial occlusion radius and a learnable radius offset, and the radii are optimized via finite differences.
This formulation enables end-to-end optimization of occlusion while maintaining training efficiency.

In summary, our contributions include:
\begin{itemize}
    \item We introduce a deformation prior extraction method that computes two vertex-wise priors from the template mesh and incorporates them into pose-dependent surfel attribute estimation for more faithful reconstruction.
    \item We propose a differentiable screen-space ambient occlusion formulation that optimizes body-part-specific occlusion radii through finite differences, enabling occlusion approximation to be jointly optimized with the avatar representation.
    \item Extensive experiments on AvatarReX and ActorsHQ datasets demonstrate that \textbf{ARS-Avatar} achieves competitive or improved reconstruction quality and enables realistic animation and relighting.
\end{itemize}

%% file: sec/2_relate.tex
\section{Related Work}
\input{figure-src/pipeline}
\subsection{Animatable Human Avatar Modeling}
Early methods rely on explicit meshes~\cite{alldieck2018video, alldieck2018detailed, xu2011video, burov2021dynamic} and template body models such as SMPL~\cite{loper2015smpl} and SMPL-X~\cite{pavlakos2019smplx} to reconstruct and deform avatars. Later methods introduce learnable deformation fields to capture pose-dependent deformations~\cite{habermann2023hdhumans, xiang2022dressing}. More recent NeRF-based methods~\cite{liu2021neural, zheng2022structured, wang2022arah, chen2023uv, kwon2023deliffas, li2023posevocab, zheng2023avatarrex} model complex geometry and appearance for continuous, high-fidelity reconstruction and novel pose animation, but ray sampling limits efficiency. Explicit 3DGS~\cite{kerbl20233d} has thus emerged as an effective representation for animatable avatars~\cite{kocabas2024hugs, li2024animatable, hu2024gaussianavatar, pang2024ash, qian20243dgs, hu2024gauhuman, zheng2024gps, zielonka2025drivable, li2025anigaussian, jiang2025uv}. Despite efficient rendering, modeling complex clothed humans remains challenging, as conditioning solely on the driving pose does not fully capture rich pose-dependent deformation. 
To address this, we propose a deformation prior extraction and incorporation method that computes deformation displacement and parametric-body deviation from the template mesh.
The deformation priors are further utilized as auxiliary for surfel attribute estimation and improve pose-dependent reconstruction fidelity.

\subsection{Relightable Human Avatar Modeling}
Relightable avatar modeling is an inverse rendering problem that disentangles geometry, materials, and illumination. Early works use neural networks to relight single images~\cite{zhou2019deep, sun2019single, wang2020single, pandey2021total, kanamori2018relighting, ji2022geometry}, but provide limited support for animation. Chen and Liu~\cite{chen2022relighting4d} pioneered integrating physically-based rendering~\cite{walter2007microfacet} into NeRF for human geometry and materials. Subsequent methods~\cite{iqbal2023rana, xiao2024neca} improve appearance modeling for detailed relighting, while others~\cite{lin2024relightable, xu2024relightable, carbonera2024relightable, wang2024intrinsic} use explicit light transport for accurate intrinsic decomposition and physically plausible relighting. Recent Gaussian avatars combine explicit primitives with physically based material and illumination models~\cite{li2023animatable, saito2024relightable, zhan2025interactive, wang2025relightable, jiang2025dnf, zeng2025rega, choi2025relightable}, 
enabling efficient and high-fidelity relighting.
We follow these works and explore Gaussian Surfels~\cite{dai2024high} as the representation primitive, leveraging their surface-aligned advantage for high-fidelity appearance and geometry.

\subsection{Visibility Approximation for Avatars}
Light visibility is crucial in physically-based rendering, determining light transport between light and surface points. The common approach computes visibility via ray tracing on meshes~\cite{chen2024meshavatar}, which typically requires high-quality meshes. Furthermore, NeRF-based methods encode visibility into MLPs~\cite{xiao2024neca, xu2024relightable}, and Lin~\etal~\cite{lin2024relightable} divide the body into parts to better model self-occlusion. Recently, visibility for explicit Gaussians has gained attention~\cite{gao2024relightable}. One strategy combines Gaussians with a mesh and uses mesh visibility~\cite{zhan2025interactive}. Liang~\etal~\cite{liang2024gs} model visibility using ambient occlusion (AO) and encode it into spherical harmonics for efficient querying, later extended to dynamic avatars by Jiang~\etal~\cite{jiang2025dnf}. Chen~\etal~\cite{chen2024gi} adopt screen-space ambient occlusion (SSAO) to improve efficiency of occlusion computation. 
However, these approaches either precompute visibility or treat it as a separate module, which makes it difficult to jointly optimize visibility with the avatar representation. In contrast, our method formulates SSAO with respect to the occlusion radius through finite differences, enabling end-to-end occlusion optimization while maintaining the efficiency required for practical avatar reconstruction.

%% file: figure-src/pipeline.tex
\begin{figure*}[t]
\centering
\includegraphics[width=0.9\textwidth]{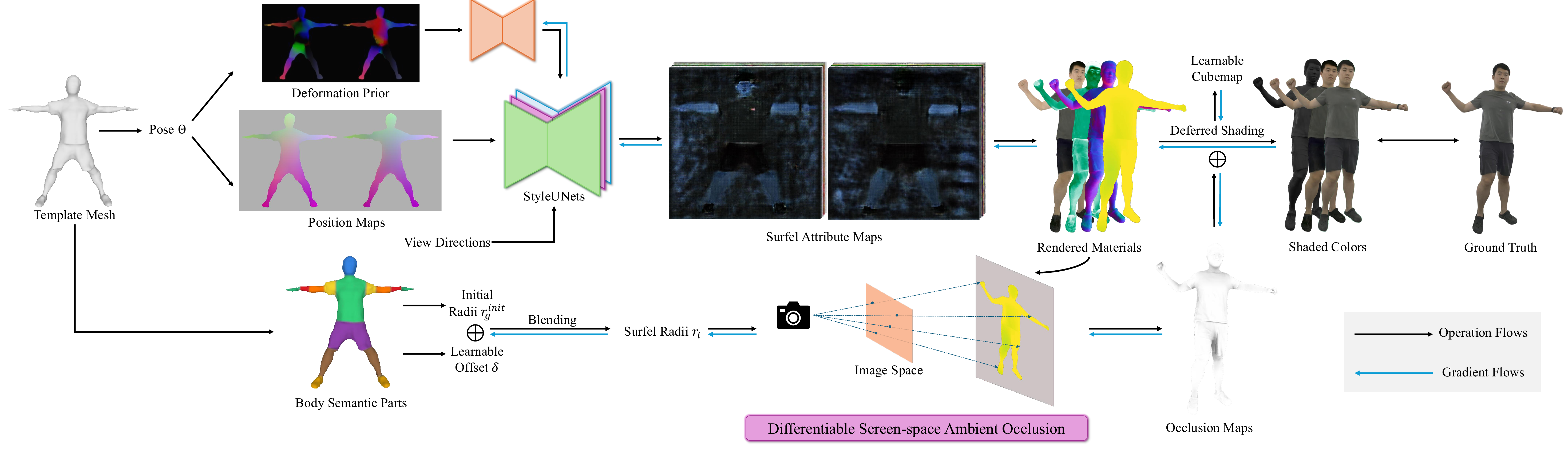}
\caption{\textbf{An illustration of our method.}
Given a canonical template mesh, we first deform it into the observation space under a driving pose $\Theta$ to generate position maps and extract deformation prior maps based on deformation priors.
Both are fed into multiple StyleUNets to estimate attributes of each surfel.
We then employ deferred shading to estimate material properties for relighting under novel illumination. 
Finally, we introduce a differentiable screen-space ambient occlusion (SSAO) module that efficiently estimates occlusion and enables joint optimization with the avatar representation during training.}
\label{fig:pipeline}
\end{figure*}

%% file: sec/3_method.tex
\section{Method}
\subsection{Overview}
Given multi-view images of a clothed human, we aim to create an animatable and relightable avatar. 
As shown in Fig.~\ref{fig:pipeline}, we represent the avatar with Gaussian Surfels~\cite{dai2024high}.
Following~\cite{li2024animatable}, we estimate pose-dependent surfel attributes from position maps using StyleUNets~\cite{wang2023styleavatar}.
To better capture the rich deformation features, we extract deformation priors from the template and incorporate them into estimation (Sec.~\ref{sec:attention}).
We further employ deferred shading to estimate BRDF materials (Sec.~\ref{sec:relight}). 
Light visibility is approximated as screen-space ambient occlusion (SSAO), made differentiable via a finite-difference formulation with respect to the occlusion radius, enabling efficient radius optimization while incorporating visibility into training (Sec.~\ref{sec:occlusion}).
Finally, \textbf{ARS-Avatar} creates animatable and relightable human avatars under any target pose, viewpoint, and illumination.

\subsection{Surfel Estimation with Deformation Priors}
\label{sec:attention}
We initialize the canonical avatar with surfels $\mathcal{G}$, each parameterized by a position $\boldsymbol{\mu}_c\in\mathbb{R}^3$, 
rotation $\mathbf{r}_c\in\mathbb{R}^{4}$,
scale $\mathbf{s}\in\mathbb{R}^3$,
opacity $\mathbf{o}\in\mathbb{R}^1$, 
and directional color $\mathbf{c}\in\mathbb{R}^3$. 
For physically-based shading, we additionally define albedo $\mathbf{a}\in\mathbb{R}^3$ and roughness $\boldsymbol{\rho}\in\mathbb{R}^1$.
Given a coarse template mesh and driving pose $\Theta$, we follow Li~\etal~\cite{li2024animatable} and rasterize template vertices $x_c$ colored by posed coordinates $x_o$ from front and back views to obtain position maps:
\begin{equation}
\mathbf{M}_{pos}^f\left(\Theta\right), \mathbf{M}_{pos}^b\left(\Theta\right) = \mathrm{Raster}\left(x_c, x_o; \Theta\right),
\end{equation}
where $\mathrm{Raster}(\cdot;\Theta)$ rasterizes the template mesh under pose $\Theta$ from front and back views.
For deformation priors, we use two vertex-wise priors: a) the distance before and after deformation $\mathbf{d}$, and b) the distance to the nearest SMPL-X~\cite{pavlakos2019smplx} vertex $\mathbf{s}$,
both positively correlated with deformation complexity. We render them similarly to obtain deformation prior maps:
\begin{equation}
\mathbf{M}_{def}^f\left(\Theta\right),\mathbf{M}_{def}^b\left(\Theta\right)=\mathrm{Raster}\left(\mathbf{x}_c,\left\{\mathbf{d},\mathbf{s}\right\};\Theta\right),
\end{equation}
where $\{\cdot\}$ denotes concatenate operation and we discuss it in the appendix.
We encode these maps with a neural network to generate deformation feature maps,
\begin{equation}
f=\mathcal{F}\left(\mathbf{M}_{def}^f\left(\Theta\right),\mathbf{M}_{def}^b\left(\Theta\right)\right),
\end{equation}
where $\mathcal{F}(\cdot)$ is the neural network.
The feature maps are further used as auxiliary to incorporate with position maps into StyleUNets to estimate surfel attributes,
\begin{equation}
\boldsymbol{\mu}_c,\mathbf{r}_c,\mathbf{s},\mathbf{o},\mathbf{c},\mathbf{a},\boldsymbol{\rho}=\mathcal{S}\left(\mathbf{M}_{pos}^f, \mathbf{M}_{pos}^b;f,\mathcal{V}\right),
\end{equation}
where $\mathcal{V}$ denotes view feature map employed in~\cite{li2024animatable} and $\mathcal{S}\left(\cdot\right)$ is the StyleUNets. We further deform canonical surfels to observation space via Linear Blend Skinning (LBS),
\begin{equation}
\boldsymbol{\mu}_o = \mathcal{R}\boldsymbol{\mu}_c + \mathcal{T},\quad \mathbf{r}_o = \mathcal{R}\mathbf{r}_c,
\end{equation}
where $\mathcal{R}$ and $\mathcal{T}$ are rotation and translation from skinning weights.
The covariance matrix is:
\begin{equation}
\boldsymbol{\Sigma}_o=\mathbf{r}_o\mathbf{s}\mathbf{s}\mathbf{r}_o^\top,
\end{equation}
finally, we render the avatar to obtain the rendered color $\hat{C}$, depth $\hat{D}$, normal $\hat{N}$, albedo $\hat{A}$, roughness $\hat{H}$, and position $\hat{P}$,
\begin{equation}
\label{eq:splat}
\hat{C},\hat{D},\hat{N},\hat{A},\hat{H},\hat{P}=\mathrm{Splat}\left(\mathbf{c},\mathbf{a},\boldsymbol{\rho};\boldsymbol{\mu}_o,\boldsymbol{\Sigma}_o,\mathbf{o}\right),
\end{equation}
where $\mathrm{Splat}(\cdot;\cdot)$ denotes EWA splatting~\cite{zwicker2001ewa} adapted from~\cite{dai2024high} and extended to render additional feature maps. We further extend the formulation in Sec.~\ref{sec:occlusion}.

\subsection{Material Estimation with Deferred Shading}
\label{sec:relight}
We follow the rendering equation~\cite{rendering2015physically} for physically-based shading,
\begin{equation}
L_o(\omega_o)=\int_{\Omega}L_i(\omega_i)f_r(\omega_i,\omega_o)(\omega_i\cdot\mathbf{n})d\omega_i,
\end{equation}
where $L_o(\omega_o)$ and $L_i(\omega_i)$ are outgoing radiance in
direction $\omega_o=P_{\mathrm{cam}}-\hat{P}$ and incident light from $\omega_i$.
The incident light is represented by a learnable HDR cube map with the shape $6\times512\times512$.
We use the Disney shading model~\cite{burley2012physically} with albedo $\mathbf{a}$, and roughness $\boldsymbol{\rho}$ as BRDF materials.
We decompose the BRDF term $f_r$ into diffuse $f_{\mathrm{diff}}$ and specular $f_{\mathrm{spec}}$. The shaded color is then computed as, 
\begin{equation}
\hat{C}_{\mathrm{shade}} = C_{\mathrm{diff}} + C_{\mathrm{spec}},
\end{equation}
we discuss the definitions of these two terms in the appendix.
Combining with ambient occlusion $\mathcal{O}$ discussed in Sec.~\ref{sec:occlusion}, the physically-based shading color is:
\begin{equation}
\hat{C} = \mathcal{O} \odot \hat{C}_{\mathrm{shade}},
\end{equation}
where $\odot$ denotes element-wise multiplication. 
The learnable cube map $L_i$ is optimized jointly with surfels to estimate incident illumination from observations.

\subsection{Differentiable Screen-space Ambient Occlusion}
\label{sec:occlusion}
We approximate light visibility with screen-space ambient occlusion (SSAO)~\cite{bavoil2008screen}, which calculates local visibility from the rendered depth $\hat{D}$ and normal $\hat{N}$. For each pixel $p$,
\begin{equation}
\mathrm{SSAO}(p) = \frac{1}{N}\cdot\sum_{i=1}^{N} V(p,\omega_i),
\end{equation}
where $V(p, \omega_i)\in[0,1]$ is visibility along sampled direction $\omega_i$.
However, standard SSAO is not directly amenable to end-to-end optimization, and optimizing its input depth and normal maps would implicitly modify the underlying geometry and surface orientation.

We therefore introduce a dedicated optimizable path that only adjusts occlusion radius while keeping geometry inputs fixed.
We divide the human body into nine semantic parts,
each with an initial radius and a learnable scalar offset. For part $g$:
\begin{equation}
r_g = r_g^{\mathrm{init}}\left(1+\delta_g\right), \quad g=1,\ldots,9,
\end{equation}
the radius associated with each surfel is obtained by blending the radii of its neighboring joints according to the LBS weights:
\begin{equation}
r_i=\sum_{j=1}^{J}w_{ij} r_{g(j)},
\end{equation}
where $w_{ij}$ denotes the LBS weight of surfel $i$ for joint $j$, and $g(j)$ maps joint $j$ to its corresponding body part.

We generate perturbed radii $r_g^{+}$ and $r_g^{-}$ by adding a perturbation offset $\boldsymbol{\varepsilon}\Delta_g$ to the corresponding body parts, where $\Delta_g\in\left\{-1,+1\right\}$ and $\boldsymbol{\varepsilon}$ is a small perturbation magnitude.
We splat above $3$ radii to obtain radius map $\hat{R}, \hat{R}_+, \hat{R}_-$ via Eq.~\ref{eq:splat}.
The final and perturbed occlusion map is computed as:
\begin{equation}
\hat{O},\hat{O}_{+},\hat{O}_{-}=\mathcal{A}\left(\hat{N},\hat{D};\hat{R},\hat{R}_{+},\hat{R}_{-}\right),
\end{equation}
where $\mathcal{A}$ denotes SSAO generation function.
Therefore, we estimate the gradient via the finite difference formulation,
\begin{equation}
\frac{\partial\hat{O}}{\partial \hat{R}}\approx\frac{\hat{O}_{+}-\hat{O}_{-}}{\hat{R}_{+}-\hat{R}_{-}},
\end{equation}
since PyTorch~\cite{paszke2019pytorch} and the EWA splatting~\cite{zwicker2001ewa} are differentiable, the full gradient for radius compensation is:
\begin{equation}
\frac{\partial\mathcal{L}}{\partial\delta}=\frac{\partial\mathcal{L}}{\partial\hat{O}}\cdot\frac{\partial\hat{O}}{\partial\hat{R}}\cdot\frac{\partial\hat{R}}{\partial r_i}\cdot\frac{\partial r_i}{\partial\delta_g},
\end{equation}
finally, the final shading result is obtained by:
\begin{equation}
\hat{C}_{\mathrm{pbr}}=C_{\mathrm{shade}}\cdot\hat{O},
\end{equation}
we provide a more detailed explanation in the appendix.

\subsection{Loss Function and Optimization} 
We optimize ARS-Avatar using multi-view RGB images $C$.
The training losses combines L1, perceptual, geometry, normal deformation, light, BRDF, and regularization terms:
\begin{equation}
\mathcal{L} = \mathcal{L}_{1}+\mathcal{L}_{\mathrm{lpips}}+\mathcal{L}_{\mathrm{geo}}+\mathcal{L}_{\mathrm{nd}}+\mathcal{L}_{\mathrm{reg}} + \mathcal{L}_{\mathrm{TV}} + \mathcal{L}_{\mathrm{light}},
\end{equation}
where the geometry loss enforces consistency between rendered normal and depth map. We further propose $\mathcal{L}_{\mathrm{nd}}$ to regularize normal deformation across neighboring surfels,
\begin{equation}
\mathcal{L}_{\mathrm{nd}}=\frac{1}{|\mathcal{K}|}\sum_{(i,j)\in\mathcal{K}}w_{ij}\left(1-\mathbf{n}_{oi}\cdot\mathbf{n}_{oj}\right),
\end{equation}
where $\mathbf{n}_{oi}$ denotes the observation normal of $i$-th surfel, and $\mathcal{K}$ contains selected neighboring pairs satisfying the canonical distance constraint. 
$w_{ij}$ is a spatial weight based on observation space distance between surfels $i$ and $j$, which downweights distant pairs. Its exact form and definition of the local scale $\sigma_i$ are provided in the appendix.

%% file: sec/4_experiment.tex
\section{Experiments}
\input{figure-src/quantitive_results}
\input{figure-src/comparison_radiance}
\input{figure-src/comparison_relight}
\input{figure-src/ablation_occlusion}
\noindent\textbf{Dataset.}
The experiments are conducted on AvatarReX~\cite{zheng2023avatarrex} and ActorsHQ~\cite{icsik2023humanrf} datasets.
We follow the setting in \cite{li2024animatable} and select $4$ characters from AvatarReX and $2$ actors from ActorsHQ dataset for training.

\noindent\textbf{Metric.}
We employ Peak Signal-to-Noise Ratio (PSNR), Structural Similarity Index Measure (SSIM)~\cite{wang2004image}, Learned Perceptual Image Patch Similarity (LPIPS)~\cite{zhang2018unreasonable}, and Frechet Inception Distance (FID)~\cite{heusel2017gans} as metrics.

\noindent\textbf{Baseline.}
We compare radiance reconstruction against 3DGS-Avatar~\cite{qian20243dgs}, WebAvatar~\cite{zhan2026high}, and AnimatableGaussian~\cite{li2024animatable}, and further compare physically-based shading and material estimation against MeshAvatar~\cite{chen2024meshavatar} and Relighting4D~\cite{chen2022relighting4d}.

\noindent\textbf{Results.}
We report animation and relighting results of clothed humans created by ARS-Avatar in Fig.~\ref{fig:teaser} and Fig.~\ref{fig:quantitive_results}.
\input{figure-src/ablation_attention}
\input{figure-src/ablation_surfel}


\subsection{Comparison}
\noindent\textbf{Animatable Avatar Reconstruction.}
We first evaluate the animation capability of ARS-Avatar against the evaluated avatar reconstruction methods, including 
3DGS-Avatar~\cite{qian20243dgs}, AnimatableGaussians~\cite{li2023animatable}, and WebAvatar~\cite{zhan2026high}. Qualitative and quantitative results are reported in Fig.~\ref{fig:comparison_radiance} and Tab.~\ref{tab:radiance_comparison}.
ARS-Avatar achieves competitive metric results and high-fidelity visual results under novel poses, especially in regions with complex nonrigid deformation, such as loose garments and articulated body parts.
In contrast, existing methods generally rely on rigid deformation and driving pose information, which may fail to capture rich pose-dependent details of clothed humans.
\input{tables/radiance_comparison}
\input{tables/physical_comparison}
\input{tables/ablation_attention}

\noindent\textbf{Relightable Avatar Reconstruction.}
We further compare the relighting capability of RAS-Avatar with recent physically-based avatar reconstruction methods, including MeshAvatar~\cite{chen2024meshavatar}, and Relighting4D~\cite{chen2022relighting4d}. Qualitative and quantitative comparisons are reported in Fig.~\ref{fig:comparison_relight} and Tab.~\ref{tab:physical_comparison}, respectively.
ARS-Avatar produces more consistent shading and occlusion effects than the evaluated PBR baselines under novel lighting.
Although explicit mesh or neural field methods achieve physically based appearance modeling, their visibility estimation relies on explicit geometry processing or expensive light transport. In contrast, Our introduced learnable screen-space ambient occlusion formulation enables efficient visibility approximation and joint optimization with the avatar. This yields more consistent occlusion and contact visibility without costly ray tracing or offline computation.

\subsection{Ablation Study}
We conduct extensive ablation studies to evaluate the effectiveness of our proposed components.

\noindent\textbf{Effect of Deformation Priors.}
We compare the model without deformation priors with our full model, where pose-dependent surfel attributes are directly predicted from the position map.
As shown in Fig.~\ref{fig:ab_attention}, 
introducing deformation priors improves all evaluated metrics
by providing additional spatial guidance for learning pose-dependent offsets. This enables the model to better preserve fine-grained pose-dependent details, especially in regions with large deformations, such as the limbs.

\noindent\textbf{Effect of Differentiable Screen-space Ambient Occlusion.}
We validate the results and time costs with five variants in Fig.~\ref{fig:ab_occlusion}.
The overhead introduced by the learnable radius is negligible, as the addition operations involving the radius account for only a small fraction of the total time.
Compared with RAGA-style occlusion, which is post-computed and cannot be jointly optimized with the avatar, our method produces more accurate occlusion maps, especially in self-occluded regions. As shown in Fig.~\ref{fig:ab_occlusion}, compared with mesh-based AO, which requires additional mesh extraction and ray tracing, our method achieves comparable occlusion accuracy with significantly less time. Therefore, considering the improved occlusion accuracy and the ability to jointly optimize visibility with the avatar, the overhead of our differentiable SSAO is acceptable.

\noindent\textbf{Effect of Gaussian Surfel Representation.}
Unlike conventional 3D Gaussian primitives, Gaussian Surfels explicitly encode surface orientation via surface-aligned anisotropic primitives, providing more reliable geometry for physically-based rendering. As shown in Fig.~\ref{fig:ab_surfel}, they produce more coherent depth and normal maps with sharper geometric boundaries than conventional Gaussians, which benefit material estimation and visibility modeling and leads to more realistic relighting.

%% file: figure-src/quantitive_results.tex
\begin{figure*}[!htbp]
\centering
\includegraphics[width=0.78\textwidth]{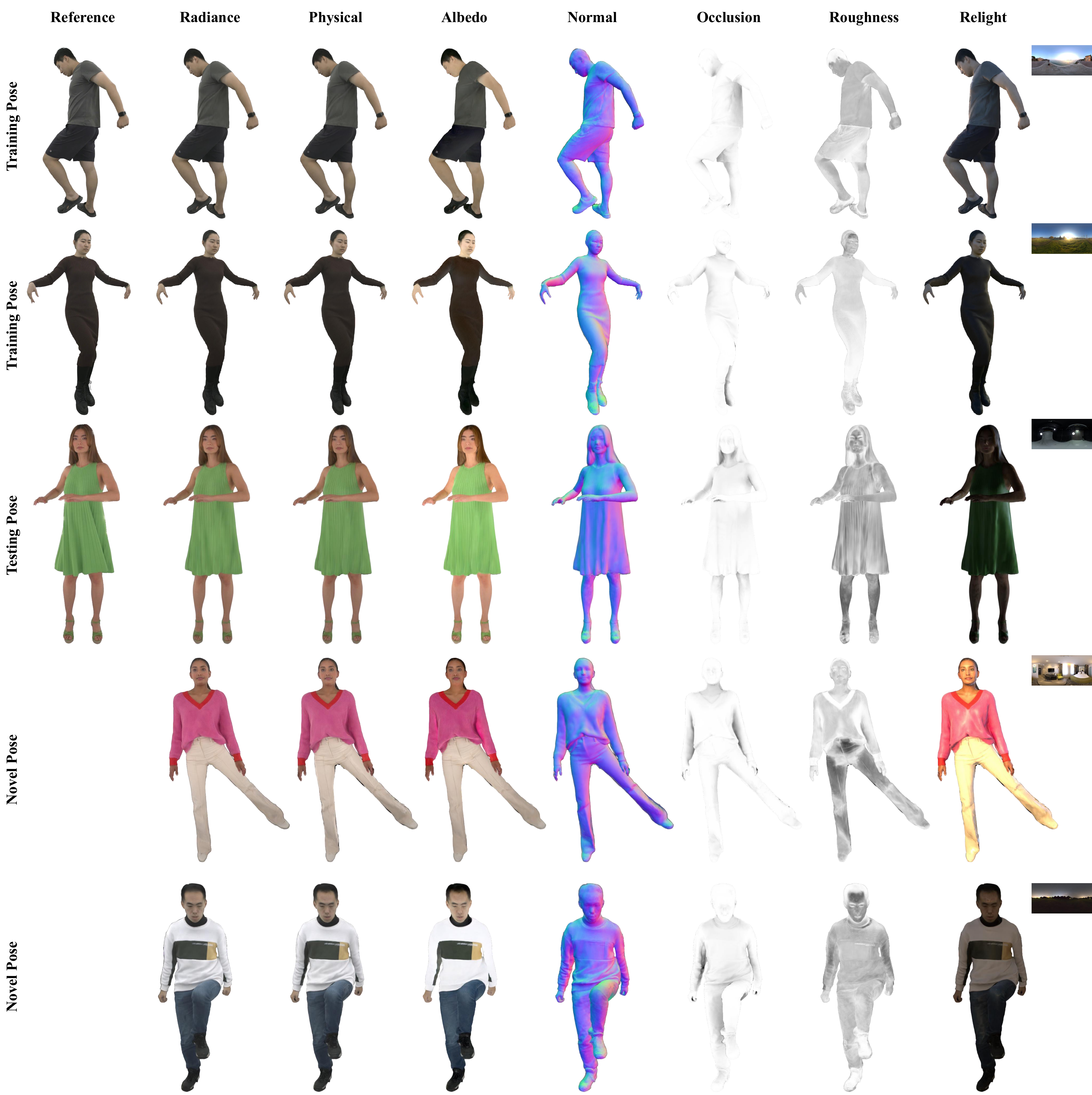}
\caption{\textbf{Illustration of appearance and geometry reconstruction, BRDF materials estimation, and relighting synthesis under training and novel poses.} The second column shows the results of directly fitting Surfel colors, while the third column shows the results obtained using physically-based rendering.}
\label{fig:quantitive_results}
\end{figure*}

%% file: figure-src/comparison_radiance.tex
\begin{figure*}[!htbp]
\centering
\includegraphics[width=0.8\textwidth]{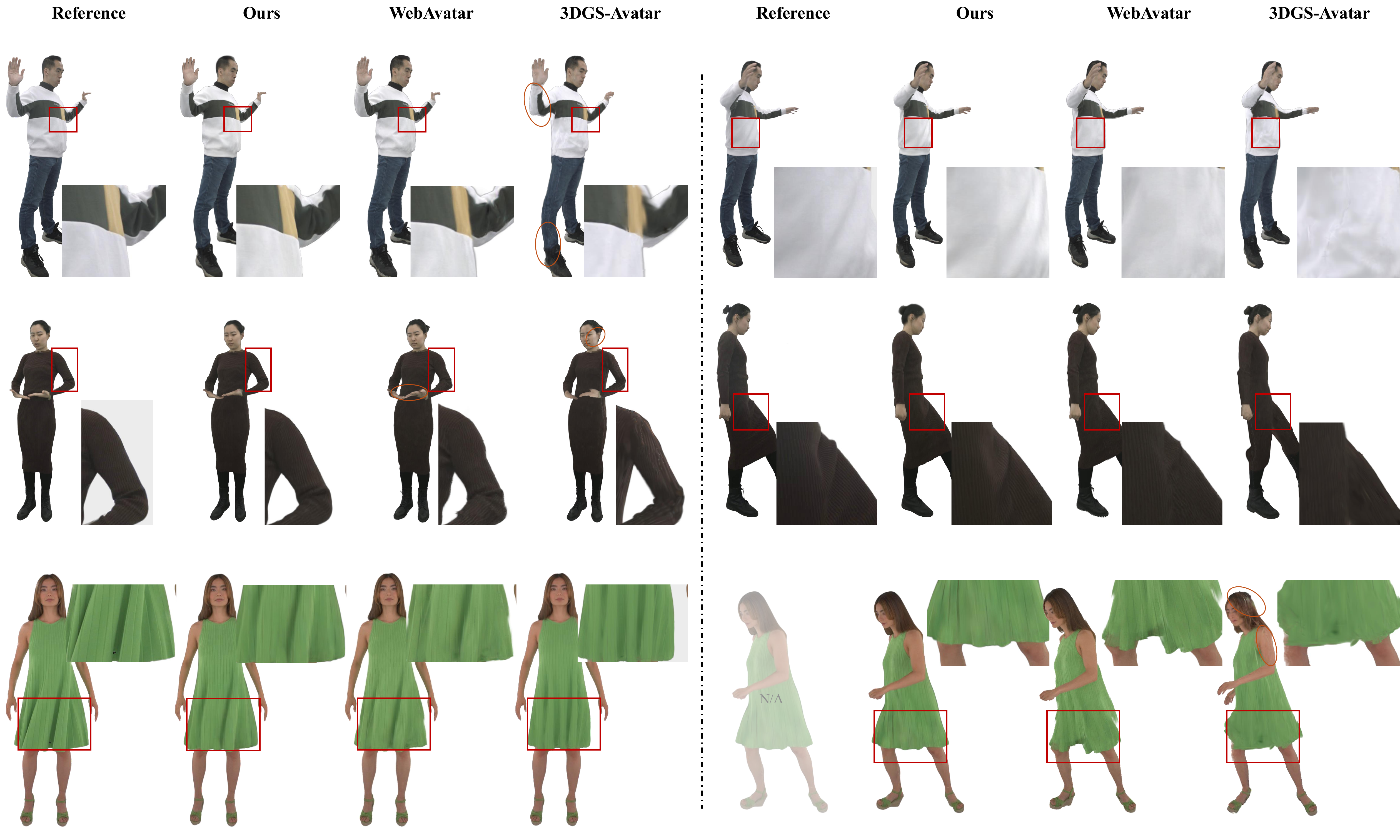}
\caption{\textbf{Qualitative Radiance Comparison with evaluated human avatar reconstruction methods including 3DGS-Avatar~\cite{qian20243dgs} and WebAvatar~\cite{zhan2026high} under testing and novel poses.}}
\label{fig:comparison_radiance}
\end{figure*}

%% file: figure-src/comparison_relight.tex
\begin{figure*}[!htbp]
\centering
\includegraphics[width=0.85\textwidth]{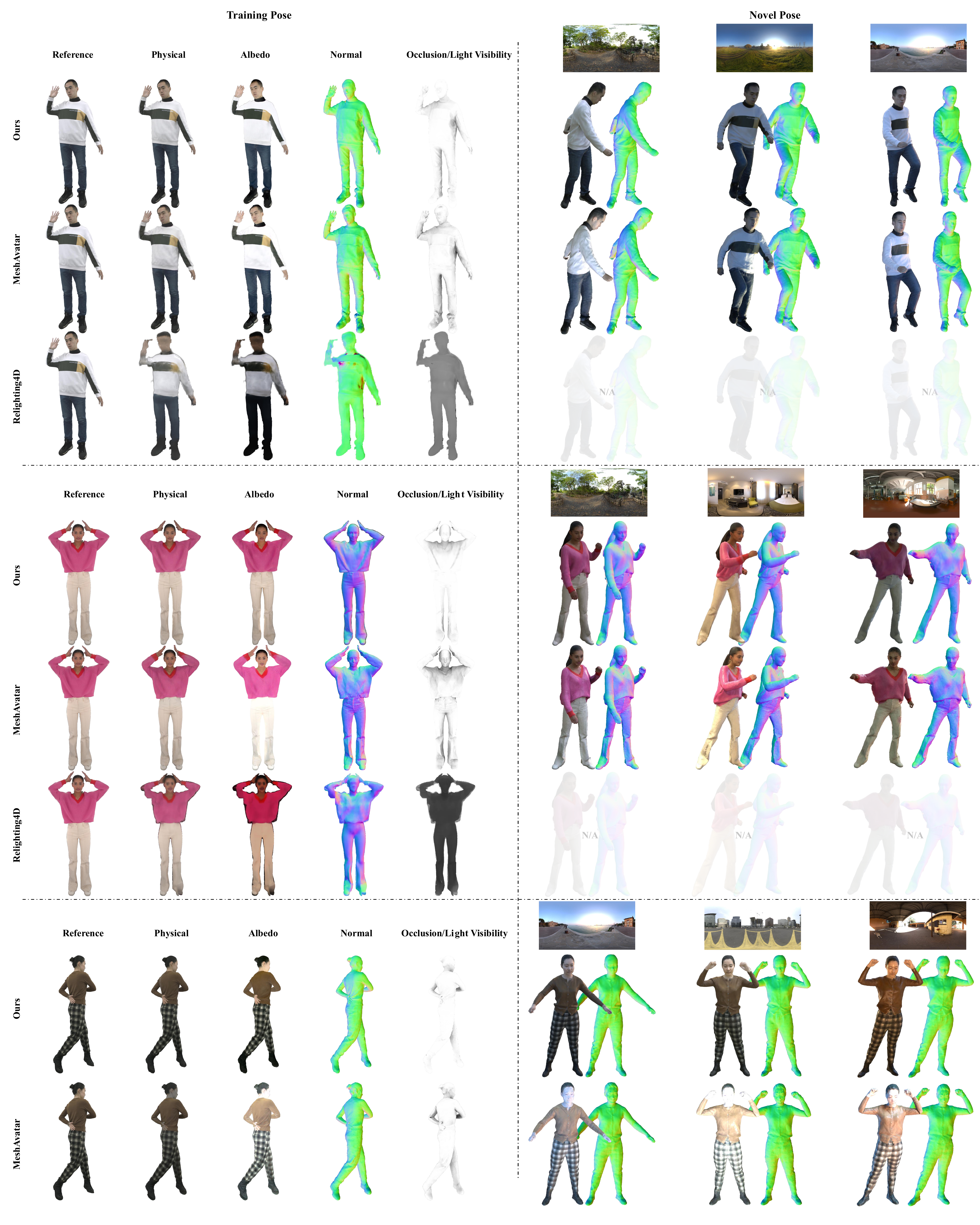}
\caption{\textbf{Qualitative Physical Comparison with evaluated human avatar relighting methods including meshavatar~\cite{chen2024meshavatar} and relighting4D~\cite{chen2022relighting4d} under testing and novel poses.}}
\label{fig:comparison_relight}
\end{figure*}

%% file: figure-src/ablation_occlusion.tex
\begin{figure*}[!htbp]
\centering
\includegraphics[width=0.85\textwidth]{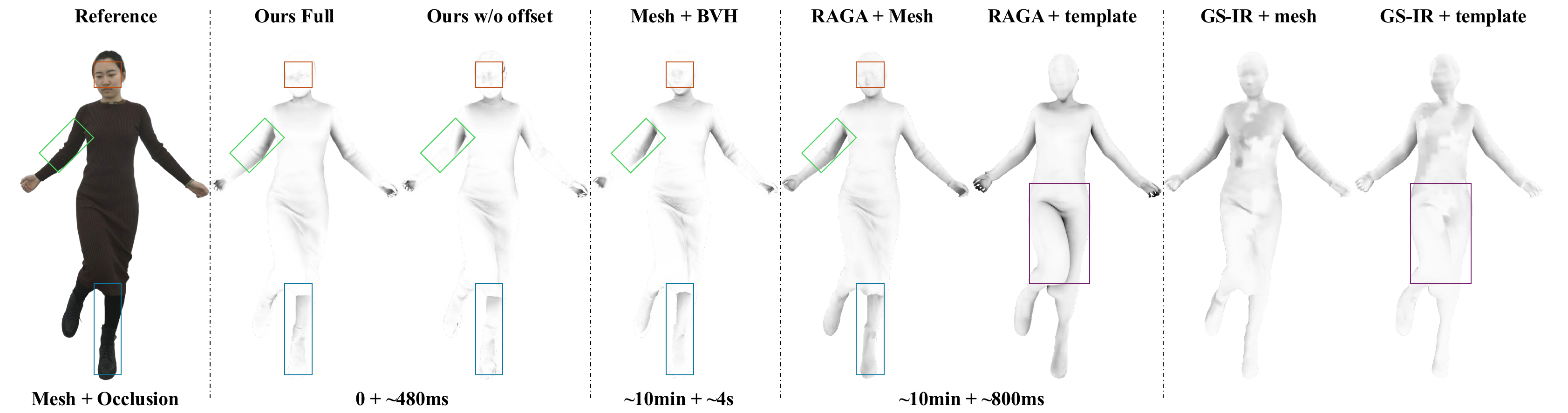}
\caption{\textbf{Qualitative Ablation on Differentiable Screen-space Ambient Occlusion.}
We compare five variants: (1) our full method; (2) our method without the optimizable radii offsets; (3) mesh-based~\cite{dai2024high} occlusion accelerated by BVH~\cite{kay1986ray}; (4) RAGA style occlusion~\cite{zhan2025interactive}; (5) GS-IR style occlusion~\cite{liang2024gs}. Our method produces high-quality occlusion results in real time without additional operation. Time for mesh extraction and occlusion computation is reported below. Variant (5) is omitted from the time comparison due to unstable results.}
\label{fig:ab_occlusion}
\end{figure*}

%% file: figure-src/ablation_attention.tex
\begin{figure}[!htbp]
\centering
\includegraphics[width=0.4\textwidth]{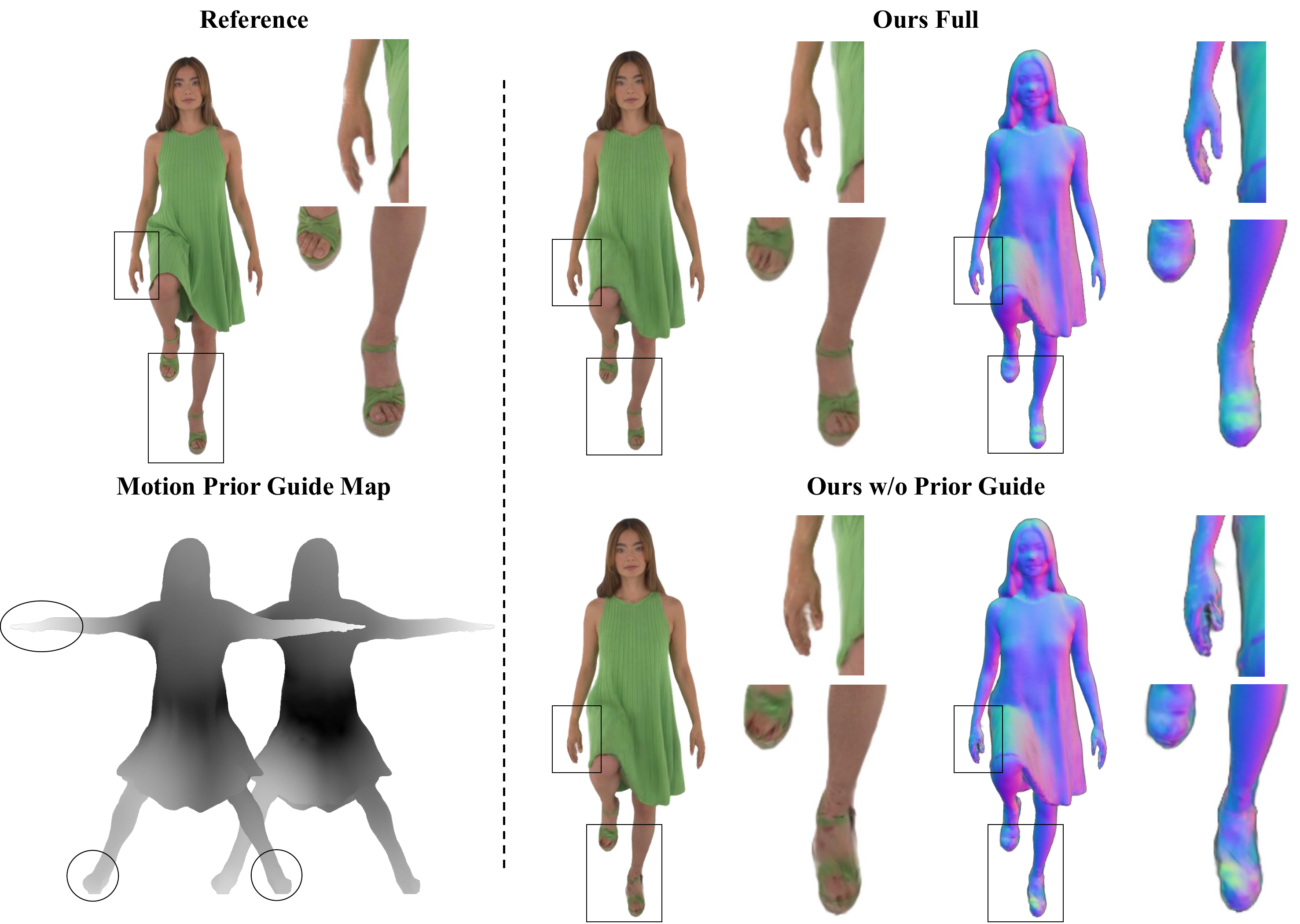}
\caption{\textbf{Qualitative Ablation on Deformation Priors.} Our deformation prior focuses on regions with significant deformation, such as the hands and feet. Incorporating it into surfel estimation yields more refined reconstructions in these regions.}
\label{fig:ab_attention}
\end{figure}

%% file: figure-src/ablation_surfel.tex
\begin{figure}[!htbp]
\centering
\includegraphics[width=0.4\textwidth]{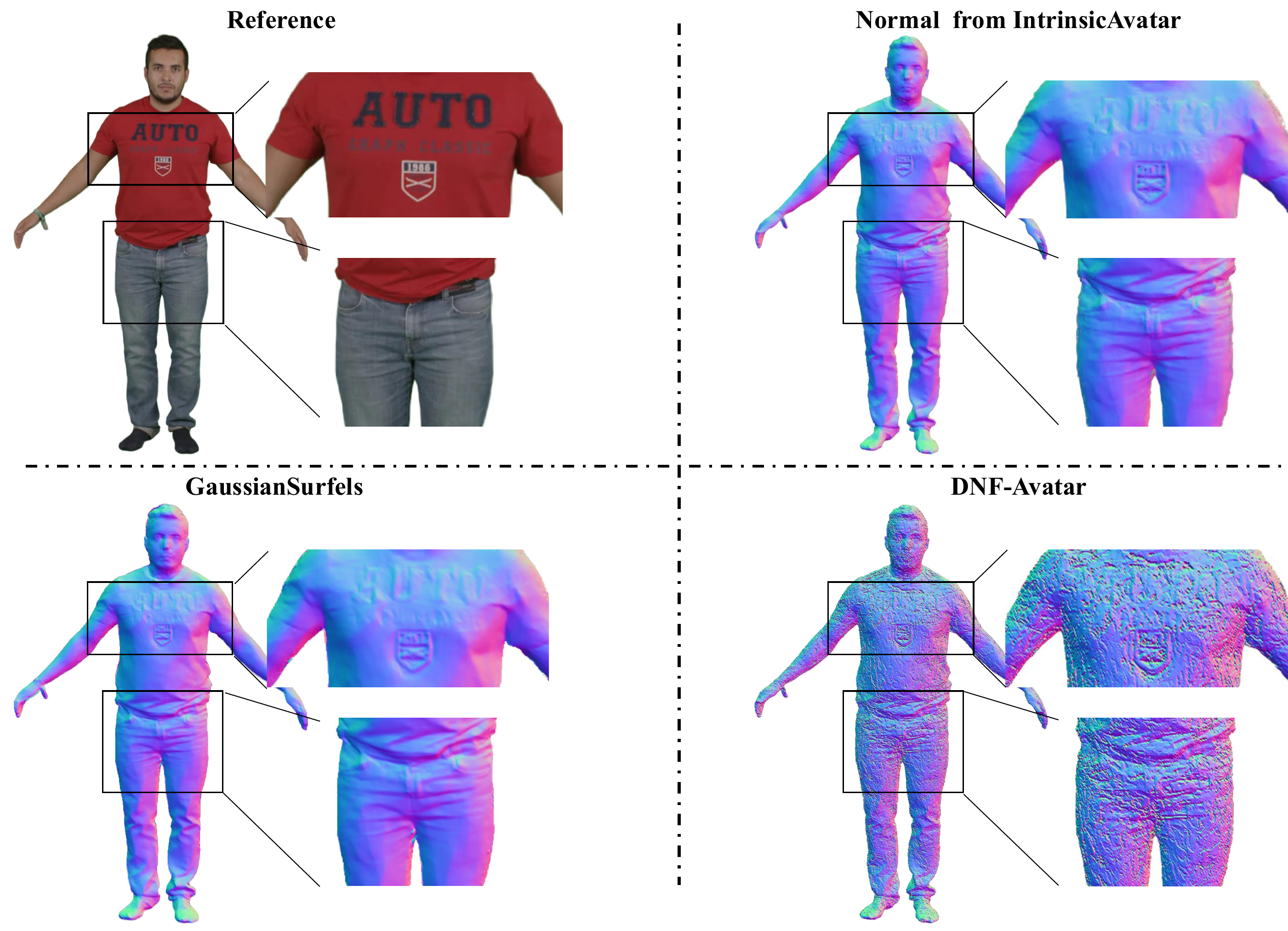}
\caption{\textbf{Qualitative Ablation on Surfel Representation.} Compared with \cite{huang20242d}, gaussiansurfel~\cite{dai2024high} representation reconstructs more accurate depth. We convert the depth into normal for clearer comparison. Training process is based on~\cite{jiang2025dnf} and normal supervision is derived from~\cite{wang2024intrinsic}.}
\label{fig:ab_surfel}
\end{figure}

%% file: tables/radiance_comparison.tex
\begin{table}[!htbp]
\caption{\textbf{Quantitative Comparison against state-of-the-art radiance reconstruction baseline on AvatarReX~\cite{zheng2023avatarrex} and ActorsHQ~\cite{icsik2023humanrf} dataset.} All methods are evaluated under the same settings. $*$Results of AG~\cite{li2024animatable} are from their paper.}
\label{tab:radiance_comparison}
\centering
\footnotesize
\begin{tabular}{cc|cccc}
\toprule
\multicolumn{1}{c}{} & \multicolumn{1}{c | }{Method} & PSNR~$\uparrow$ & SSIM~$\uparrow$ & LPIPS~$\downarrow$ & FID~$\downarrow$ \\
\midrule
\multirow{3}{*}{\rotatebox{90}{\textbf{REX}}} & 3DA~\cite{qian20243dgs} & 29.2612 & 0.9729 & 0.0389 & 29.6184 \\
& AG$^*$~\cite{li2024animatable} & 30.6143 & 0.9803 & \textbf{0.0290} & \underline{13.2417} \\
& WEB~\cite{zhan2026high} & \textbf{31.6588} & \underline{0.9809} & \underline{0.0297} & 23.1407 \\
\cmidrule{2-6}
& \textbf{Ours} & \underline{30.7514} & \textbf{0.9862} & 0.0318 & \textbf{13.1050}\\
\midrule
\multirow{3}{*}{\rotatebox{90}{\textbf{HQ}}} & 3DA~\cite{qian20243dgs} & 30.1744 & 0.9567 & \underline{0.0317} & \textbf{16.7763}\\
& AG$^*$~\cite{li2024animatable} & 30.3607 & 0.9682 & 0.0339 & 33.4665\\
& WEB~\cite{zhan2026high} & \underline{31.3856} & \underline{0.9686} & 0.0324 & 31.5130\\
\cmidrule{2-6}
& \textbf{Ours} & \textbf{31.4785} & \textbf{0.9694} & \textbf{0.0299} & \underline{18.9706}\\
\bottomrule
\end{tabular}
\end{table}

%% file: tables/physical_comparison.tex
\begin{table}[!htbp]
\caption{\textbf{Quantitative Comparison against the evaluated PBR baselines on AvatarReX~\cite{zheng2023avatarrex} and ActorsHQ~\cite{icsik2023humanrf} dataset.} ARS-Avatar outperforms the baselines on the evaluated metrics.}
\label{tab:physical_comparison}
\centering
\footnotesize
\begin{tabular}{cc|cccc}
\toprule
\multicolumn{1}{c}{} & \multicolumn{1}{c | }{Method} & PSNR~$\uparrow$ & SSIM~$\uparrow$ & LPIPS~$\downarrow$ & FID~$\downarrow$ \\
\midrule
\multirow{3}{*}{\rotatebox{90}{\textbf{REX}}} & R4D~\cite{chen2022relighting4d} & 23.6745 & 0.9252 & 0.0974 & 96.4823 \\
& MA~\cite{chen2024meshavatar} & 28.5608 & 0.9707 & 0.0377 & 18.2602 \\
\cmidrule{2-6}
& \textbf{Ours} & \textbf{30.2703} & \textbf{0.9752} & \textbf{0.0375} & \textbf{12.7732}\\
\midrule
\multirow{3}{*}{\rotatebox{90}{\textbf{HQ}}} & R4D~\cite{chen2022relighting4d} & 24.1780 & 0.9457 & 0.0627 & 139.2214 \\
& MA~\cite{chen2024meshavatar} & 28.7860 & 0.9510 & 0.0305 & 19.7140\\
\cmidrule{2-6}
& \textbf{Ours} & \textbf{31.6116} & \textbf{0.9596} & \textbf{0.0261} & \textbf{17.1087}\\
\bottomrule
\end{tabular}
\end{table}

%% file: tables/ablation_attention.tex
\begin{table}[!htbp]
\caption{\textbf{Quantitative Ablation on Deformation Priors (DP).}}
\label{tab:ablation_attention}
\centering
\footnotesize
\begin{tabular}{c|cccc}
\toprule
\multicolumn{1}{c | }{Method} & PSNR~$\uparrow$ & SSIM~$\uparrow$ & LPIPS~$\downarrow$ & FID~$\downarrow$ \\
\midrule
Ours w/o MPG & 30.9457 & 0.9515 & 0.0446 & 34.5386 \\
\midrule
\textbf{Ours} & \textbf{31.7459} & \textbf{0.9560} & \textbf{0.0382} & \textbf{33.4016} \\
\bottomrule
\end{tabular}
\end{table}

%% file: sec/5_discuss.tex
\section{Conclusion}
We present \textbf{ARS-Avatar}, which creates animatable and relightable human avatars from multi-view images. Surfel primitives achieve high fidelity and consistency in appearance and geometry with faster rendering than implicit methods. We further propose a differentiable ambient occlusion method for shadow estimation in the 3DGS pipeline, which can be integrated into optimization without additional network components while maintaining efficiency. 
Our method offers an effective solution for animatable and relightable 3D human avatar modeling.

However, It is undeniable that the network requires a long time to complete optimization, and our deformation prior extraction relies on predefined priors, which not capture diverse information across clothed human body. Future work could incorporate richer physical knowledge or priors from large-language model~\cite{liu2024deepseek, achiam2023gpt}. 
Moreover, while our current SSAO only optimizes body-part-specific occlusion radii, it could be integrated into StyleUNets so that all SSAO-related parameters, rather than only the radius, are estimated and jointly optimized with the avatar.
We leave this exploration for future work. 



%% file: sec/6_supp.tex
\section{Implementation Details}
\label{sec:impl}
All models are implemented in PyTorch and trained on a single NVIDIA RTX~3090 GPU with 24~GB of memory. 
The training iterations for radiance reconstruction and physically-based shading estimation are set to 500k and 300k, with training times of approximately $4$ and $2$ days, respectively.
The rasteriser, the cubemap filtering and the differentiable ambient occlusion formulation are custom CUDA extensions. The split-sum pre-filtering uses the same cosine and GGX kernels as the public \texttt{nvdiffrec} release. The PCA operation for out-of-distribution pose~\cite{li2024animatable} is also performed on position map and deformation priors map for more faithful reconstruction.

\section{Detailed Deformation Feature Maps}
As mentioned in Sec.~3.2, we define the deformation feature maps by utilizing two distance: a) the distance before and after deformation $\mathbf{d}$, and b) the distance to the nearest SMPL-X~\cite{pavlakos2019smplx} vertex $\mathbf{s}$. 
For the two distances, we encode their magnitude and orientation into the feature map $\mathbf{M}_{def}^{f}$ and $\mathbf{M}_{def}^{b}$ respectively. However, for distance $\mathbf{s}$, it measures the error of the weights in the LBS deformation process. Therefore, we believe that its magnitude is more valuable. Therefore, the final encoded information includes the magnitude and orientation of $\mathbf{d}$ and the magnitude of $\mathbf{s}$,
\begin{equation}
\left\{\mathbf{d}, \mathbf{s}\right\}=\mathrm{Concate}\left(\|\mathbf{d}\|, \frac{\mathbf{d}}{\|\mathbf{d}\|},\|\mathbf{s}\|\right),
\end{equation}
where $\mathrm{Concate}$ denotes concatenate operation.

\section{Detailed Splatting Formulation}
We employ $\mathrm{Splat}(\cdot;\cdot)$ introduced in GaussianSurfel~\cite{dai2024high}, which render color $\hat{C}$, normal $\hat{N}$, depth $\hat{D}$ maps from a set of surfel. We extend the formulation in manuscript to render albedo $\hat{A}$, roughness $\hat{H}$, position $\hat{P}$, radius $\hat{R}$, perturbed radius $\hat{R}_+$, $\hat{R}_-$, sample importance $I$, and SSAO bone-aware direction $Q$ maps. Therefore, the final splatting formulation is defined as:
\begin{equation}
\begin{aligned}
\hat{C},\hat{D},\hat{N},&\hat{A},\hat{H},\hat{P},\hat{R},\\ \hat{R}_+,\hat{R}_-,I,Q =& \mathrm{Splat}\left(\mathbf{c}, \mathbf{a}, \boldsymbol{\rho};\boldsymbol{\mu}_o,\boldsymbol{\Sigma}_o,\mathbf{o}\right),    
\end{aligned}
\end{equation}
we will discuss the sample importance map $I$ and SSAO direction map $Q$ in Sec.~\ref{app:ao}. The rendered albedo and roughness map is splatted without the influence of background, which avoids training failure caused by background noise during deferred shading.
\section{Detailed Physically-Based Shading}
\label{app:pbr_shading}
Specifically, we decompose the BRDF term into diffuse and specular components:
\begin{equation}
f_r(\omega_i,\omega_o)=f_{\mathrm{diff}}(\omega_i)+f_{\mathrm{spec}}(\omega_i,\omega_o),
\end{equation}
The diffuse color is computed by:
\begin{equation}
C_{\mathrm{diff}}=\frac{\mathbf{a}}{\pi}\cdot\int_\Omega L_i(\omega_i)(\omega_i\cdot\mathbf{n})d\omega_i,
\end{equation}
For specular component, we utilize SplitSum approximation to separate the light and BRDF term in the integrals. Therefore, the specular color is computed by:
\begin{equation}
\begin{aligned}
C_{\mathrm{spec}}\approx\int_\Omega& L_i(\omega_i)D(\omega_i,\omega_o)(\omega_i,n)d\omega_i \cdot \\
\int_\Omega &f_{\mathrm{spec}}(\omega_i,\omega_o)\cdot(\omega_i\cdot n)d\omega_i
\end{aligned}
\end{equation}
where $D(\omega_i,\omega_o)$ denotes normal distribution function. We discuss the terms $D$ and $f_{\mathrm{spec}}$ in the appendix.
The diffuse and specular components are finally combined to obtain the shaded
color before accounting for visibility:
\begin{equation}
\hat{C}_{\mathrm{shade}}=C_{\mathrm{diff}}+C_{\mathrm{spec}},
\end{equation}
As mentioned in Sec.~3.3, we employ the rendering equation~\cite{rendering2015physically} on rendered maps, and the BRDF term $f_r\left(\omega_i,\omega_o\right)$ is decomposed into diffuse and specular components. The diffuse component is defined as:
\begin{equation}
\label{eq_app:diffuse}
f_{\mathrm{diff}}\left(\omega_i\right)=\left(1-m\right)\cdot\frac{\mathbf{a}}{\pi},
\end{equation}
We set metallic to 0 for all surfels, as we focus on diffuse-dominant human skin and clothing. Hair and accessories with strong specularity may require a metallic channel, which we leave for future work.
We simplify the Eq.~\ref{eq_app:diffuse} by set metallic equals $0$ to get the final diffuse BRDF term $f_{\mathrm{diff}}=\frac{\mathbf{a}}{\pi}$.

For specular component $f_{\mathrm{spec}}$, we define it using normal distribution function $D$, Fresnel term $F$, and geometry term $G$, which are calculated in~\cite{walter2007microfacet},
\begin{equation}
f_{\mathrm{spec}}\left(\omega_i,\omega_o\right)=\frac{D\cdot F\cdot G}{4\cdot\left(\omega_i\cdot n\right)\cdot\left(\omega_o\cdot n\right)},
\end{equation}
where $n$ denotes the normal of current pixel.

\section{Detailed Differentiable Ambient Occlusion}
\label{app:ao}
This section provides the detailed formulation of the differentiable SSAO module introduced in Sec.~3.4. The central objective
is to enable gradient-based optimization of the occlusion radius while preserving the efficient CUDA implementation of the original SSAO operator.

\subsection{Body-Part Specific Radius Parameterization}
We divide the human body into nine semantic parts and generate a body set:
\begin{equation}
\begin{aligned}
\mathcal{B}=&\{\mathrm{torso},\mathrm{hip},\mathrm{knee},\mathrm{foot},\\&\mathrm{head},\mathrm{shoulder},\mathrm{elbow},\mathrm{wrist},\mathrm{hand}\},
\end{aligned}
\end{equation}
each body part $b\in\mathcal{B}$ is assigned an initial radius
$r_g^{\mathrm{init}}$ and a learnable offset $\delta_g$. 
We have discussed how to obtain the radius of each surfel in Eq.~$(12\sim15)$. Finally, we assign the radius attribute $r_i$ to each surfel and make it participate in splatting to obtain the rendered radius map $\hat{R}$.

\subsection{Custom Backward Propagation}

Given an upstream gradient$\frac{\partial\mathcal{L}}{\partial\hat{O}}$,
the custom backward operation computes as,
\begin{equation}
\frac{\partial\mathcal{L}}{\partial\hat{R}}=\frac{\partial\mathcal{L}}{\partial\hat{O}}\cdot\frac{\partial\hat{O}}{\partial\hat{R}},
\end{equation}
the resulting gradient is propagated through the differentiable baseline
radius branch:
\begin{equation}
\frac{\partial\mathcal{L}}{\partial\hat{R}}\rightarrow\frac{\partial\mathcal{L}}{\partial\{r_i\}}\rightarrow\frac{\partial\mathcal{L}}{\partial\{r_g\}}\rightarrow\frac{\partial\mathcal{L}}{\partial\{\delta_g\}_{g=1}^{9}},
\end{equation}
two perturbed positive and negative radius branches are detached from the computational
graph and therefore do not receive gradients. Likewise, the normal, depth,
bone-aware direction, and importance maps are treated as fixed inputs to the
differentiable SSAO module:
\begin{equation}
\frac{\partial\hat{O}}{\partial\hat{N}}=\frac{\partial\hat{O}}{\partial\hat{D}}=\frac{\partial\hat{O}}{\partial Q}=\frac{\partial\hat{O}}{\partial I}=\mathbf{0},
\end{equation}
consequently, the proposed module provides a gradient path only for the body-part specific radius parameters.




\subsection{Bone-Aware Direction Map}
To guide the SSAO sampling direction, we introduce a bone-aware direction map $Q$ that incorporates pose-dependent skeletal information into the sampling process. For each surfel primitive $i$, we first compute its bone-aware direction by blending the joint frame axes with the corresponding LBS weights:
\begin{equation}
\label{eq:boneup}
\mathbf{q}^{\mathrm{bone}}_i=\operatorname{normalize}\left(\sum_{j=1}^{J}w_{ij}\,\mathbf{V}_{j}\,\mathbf{e}_{y}\right),
\end{equation}
where $w_{ij}$ denotes the LBS weight between surfel $i$ and joint $j$, $\mathbf{e}_{y}=(0,1,0)$ is the canonical up axis, and $\mathbf{V}_{j}=\mathbf{R}_{j}\mathbf{R}^{\mathrm{rest}\,-1}_{j}$ is the pose-dependent rotation of joint $j$, expressed as the joint's canonical-to-live rotation $\mathbf{R}_{j}$ composed with the inverse of its rotation $\mathbf{R}^{\mathrm{rest}}_{j}$ in the canonical pose. 

The direction $\mathbf{q}^{\mathrm{bone}}_i$ is produced in world space and is rotated into camera space before being splatted, so that it is expressed in the same frame as the camera-space surfel normal $\mathbf{n}_i$. Since SSAO sampling is performed around the surface tangent plane, we project the bone direction onto the tangent space of each surfel:
\begin{equation}
\label{eq:proj}
\mathbf{t}'_i=\mathbf{q}^{\mathrm{bone}}_i-(\mathbf{q}^{\mathrm{bone}}_i\cdot\mathbf{n}_i)\mathbf{n}_i,
\end{equation}

\begin{equation}
\label{eq:basis}
\mathbf{t}_i=\operatorname{normalize}(\mathbf{t}'_i),\quad\mathbf{b}_i=\mathbf{n}_i\times\mathbf{t}_i,
\end{equation}
the tangent basis is then defined as:
\begin{equation}
\mathbf{Q}_i=[\mathbf{t}_i,\mathbf{b}_i,\mathbf{n}_i],
\end{equation}
Similar to other surfel attributes, the direction field $\{\mathbf{q}^{\mathrm{bone}}_i\}$ is
splatted into image space using $\mathrm{Splat}(\cdot;\cdot)$ to obtain the bone-aware direction map $Q$.

\subsection{Sample Importance Map}
Our screen-space ambient occlusion solver marches a fixed set of hemisphere directions on a uniform
angular grid, so that every pixel receives the same number of occlusion rays. This uniform
allocation is a poor fit for an articulated human body, whose parts span more than an order of
magnitude in size: because a ray-marching step maps to a screen-space displacement proportional to
the surface depth, a density tuned to resolve occlusion on the trunk wastes samples on its wide
variation, while the same density is far too sparse to resolve the narrow occlusion features of
fingers and wrists. We therefore introduce a sample importance map $I$ that allocates SSAO sampling
density according to the local body-part scale, and is cheap to evaluate because it reuses the LBS
weights and joints already available from skinning.

For each surfel $i$ we first compute its skeletal reference position by LBS-weighted interpolation
of joint locations:
\begin{equation}
\mathbf{x}^{\mathrm{bone}}_i
=
\sum_{j=1}^{J}
w_{ij}\mathbf{p}_{j},
\end{equation}
where $\mathbf{p}_{j}$ denotes the position of the $j$-th joint in the posed frame, $w_{ij}$ is the
LBS weight of surfel $i$ at joint $j$, and $J$ is the number of joints.

A uniform distance decay would treat all joints alike, but the relevant length scale differs by an order of magnitude between, e.g., a finger joint and the pelvis. We therefore associate each joint $j$ with a nominal body-part radius $r_j$ and define the importance of a surfel as the ratio between its blended body-part scale and the per-joint scale of the reference position:
\begin{equation}
I_i = \frac{\sum_{j=1}^{J} w_{ij}\,e^{-\|\mathbf{x}_i-\mathbf{p}_j\|_2/\sigma}\,r_j}{\sum_{j=1}^{J}w_{ij}\,r_j}\in\left[0,1\right],
\end{equation}
where $\mathbf{x}_i$ is the position of surfel $i$, and $\sigma=0.15\,\mathrm{m}$ is a distance fall-off
radius. The numerator is an exponentially distance-weighted average of the per-joint radii over the
joints that actually influence the surfel, so a surfel lying on a thin structure such as a finger
inherits the small radius of its governing joint and obtains $I_i\to 1$, whereas a surfel on a thick
structure such as the torso is dominated by the large radii of the trunk joints and obtains a small
$I_i$. The exponential stencil makes the transition between neighbouring body parts smooth, so no
part segmentation is required; because it appears in both numerator and denominator, the ratio is
also independent of the absolute scale of the joint radii.

Similar to other surfel attributes, the importance values are packed into the SSAO feature channel
of the surfel and splatted using $\mathrm{Splat}(\cdot;\cdot)$.
The obtained importance map $I$ is used to adjust the angular sampling density of SSAO:
\begin{equation}
\label{eq:adp}
\delta_{\mathrm{adp}}=\frac{\delta}{1+\alpha I},
\end{equation}
where $\delta=0.0625$ is the default angular sample interval, and $\alpha=3$ is a weight that controls the influence of the importance map on the sampling density. Allocating a full unit of angular budget to the most articulated regions yields a denser step $\delta_{\mathrm{adp}}\in[\delta/4,\delta]$, i.e. up to four times more sampled directions on thin appendages, while leaving the sampling on the torso unchanged. The kernel clamps the incoming importance to $[0,1]$ before applying \eqref{eq:adp} to keep the number of sampled directions bounded in all cases.

\input{tables/loss_weight}
\subsection{Detailed Training Losses}
The training process includes two stage, In the first stage, we use the directional color and reconstruct the geometry of the human avatar. The training loss consist of radiance loss, geometry loss, normal deformation loss, and regularization.
The radiance loss includes L1 and lpips loss,
\begin{equation}
\mathcal{L}_{\mathrm{rad}}=\mathcal{L}_\mathrm{1}\left(C,\hat{C}\right)+\mathcal{L}_{\mathrm{lpips}}\left(C, \hat{C}\right),
\end{equation}
and geometry loss is defined as:
\begin{equation}
\mathcal{L}_{\mathrm{geo}}=
\mathcal{L}_{\mathrm{cons}}\left(\hat{D},\hat{N}\right)+\mathcal{L}_{\mathrm{curv}},
\end{equation}
We have discussed the definition of normal deformation loss $\mathcal{L}_{\mathrm{nd}}$ and the spatial weight in maniscript is defined as,
\begin{equation}
w_{ij}=\exp\left(-\frac{\|\boldsymbol{\mu}_{oi}-\boldsymbol{\mu}_{oj}\|_2^2}{\sigma_i^2}
\right),
\end{equation}
where $\sigma_i$ is the local distance scale from the 
farthest neighbor in the $K$-nearest neighborhood of the $i$-th surfel.
We further regularize the offset $\boldsymbol{\mu}$ between the optimized and initial surfel position,
\begin{equation}
\mathcal{L}_{\mathrm{reg}}=\left\|\Delta\boldsymbol{\mu}\right\|_2^2,
\end{equation}
in the second stage, we estimate BRDF materials and light and add extended training loss, including shading loss, TV smooth loss, and regularization of light.
\begin{equation}
\mathcal{L}_{\mathrm{shaded}}=\mathcal{L}_\mathrm{1}\left(C,\hat{C}_{\mathrm{pbr}}\right)+\mathcal{L}_{\mathrm{lpips}}\left(C, \hat{C}_{\mathrm{pbr}}\right),
\end{equation}
We employ the TV smooth loss $\mathcal{L}_{\mathrm{TV }}$ on rendered albedo and rendered roughness map and environment light, and enforce the light tend to white light,
\begin{equation}
\mathcal{L}_{\mathrm{light}}=\frac{1}{3}\sum_{i=1}^{3}\left\|l_i-\frac{1}{3}\sum_{j=1}^{3}l_j\right\|.
\end{equation}
we use $\mathcal{L}_{\mathrm{rad}}$ in the first stage and switch to $\mathcal{L}_{\mathrm{shaded}}$ in the second stage, while also adding $\mathcal{L}_{\mathrm{TV}}$ and $\mathcal{L}_{\mathrm{light}}$.

\section{Ethical Considerations.}
Our method can synthesize animatable and relightable human avatars, which could be misused to generate fake videos or non-consensual avatar content. We therefore emphasize that any deployment of this technology should obtain explicit consent from the depicted individuals and should include safeguards such as watermarking and deepfake detection.

%% file: tables/loss_weight.tex
\begin{table}[!htbp]
\caption{\textbf{Loss terms and weights.} The depth normal consistency weight ramps up from $0.025$ to $0.125$ during training.}
\label{tab:losses}
\centering
\small
\setlength{\tabcolsep}{4pt}
\begin{tabular}{llcc}
\toprule
Loss & Symbol & Weight & Stage \\
\midrule
Photometric L1        & $\mathcal{L}_{\text{1}}$    & $1.0$   & Rad., PBR \\
Perceptual (LPIPS)    & $\mathcal{L}_{\mathrm{lpips}}$  & $0.1$   & Rad., PBR \\
Offset regularisation & $\mathcal{L}_{\mathrm{reg}}$    & $0.005$ & Rad., PBR \\
Depth normal consistency & $\mathcal{L}_{\mathrm{cons}}$   & $0.025$ & Rad., PBR \\
Curvature smoothness  & $\mathcal{L}_{\mathrm{curv}}$   & $0.005$ & Rad., PBR \\
Normal deformation & $\mathcal{L}_{\mathrm{nd}}$   & $0.004$ & Rad. \\
BRDF TV               & $\mathcal{L}_{\mathrm{TV}}$   & $0.005$ & PBR \\
Environment prior     & $\mathcal{L}_{\mathrm{light}}$    & $0.05$  & PBR \\
\bottomrule
\end{tabular}
\end{table}

%% file: main.bib
@String(CVPR= {IEEE Conf. Comput. Vis. Pattern Recog.})

@String(TOG= {ACM Trans. Graph.})

@String(ICME = {Int. Conf. Multimedia and Expo})

@String(AAAI = {AAAI})

@String(CVPR  = {CVPR})

@String(TOG   = {ACM TOG})

@String(ICME  =	{ICME})

@article{guo2019relightables,
  title={The relightables: Volumetric performance capture of humans with realistic relighting},
  author={Guo, Kaiwen and Lincoln, Peter and Davidson, Philip and Busch, Jay and Yu, Xueming and Whalen, Matt and Harvey, Geoff and Orts-Escolano, Sergio and Pandey, Rohit and Dourgarian, Jason and others},
  journal={ACM Transactions on Graphics (ToG)},
  volume={38},
  number={6},
  pages={1--19},
  year={2019},
  publisher={ACM New York, NY, USA}
}

@inproceedings{debevec2000acquiring,
  title={Acquiring the reflectance field of a human face},
  author={Debevec, Paul and Hawkins, Tim and Tchou, Chris and Duiker, Haarm-Pieter and Sarokin, Westley and Sagar, Mark},
  booktitle={Proceedings of the 27th annual conference on Computer graphics and interactive techniques},
  pages={145--156},
  year={2000}
}

@article{mildenhall2021nerf,
  title={Nerf: Representing scenes as neural radiance fields for view synthesis},
  author={Mildenhall, Ben and Srinivasan, Pratul P and Tancik, Matthew and Barron, Jonathan T and Ramamoorthi, Ravi and Ng, Ren},
  journal={Communications of the ACM},
  volume={65},
  number={1},
  pages={99--106},
  year={2021},
  publisher={ACM New York, NY, USA}
}

@inproceedings{xu2024relightable,
  title={Relightable and animatable neural avatar from sparse-view video},
  author={Xu, Zhen and Peng, Sida and Geng, Chen and Mou, Linzhan and Yan, Zihan and Sun, Jiaming and Bao, Hujun and Zhou, Xiaowei},
  booktitle={Proceedings of the IEEE/CVF Conference on Computer Vision and Pattern Recognition},
  pages={990--1000},
  year={2024}
}

@inproceedings{lin2024relightable,
  title={Relightable and animatable neural avatars from videos},
  author={Lin, Wenbin and Zheng, Chengwei and Yong, Jun-Hai and Xu, Feng},
  booktitle={Proceedings of the AAAI Conference on Artificial Intelligence},
  volume={38},
  number={4},
  pages={3486--3494},
  year={2024}
}

@inproceedings{wu2025fast,
  title={Fast and Physically-based Neural Explicit Surface for Relightable Human Avatars},
  author={Wu, Jiacheng and Zhang, Ruiqi and Chen, Jie and Zhang, Hui},
  booktitle={2025 IEEE International Conference on Multimedia and Expo (ICME)},
  pages={1--6},
  year={2025},
  organization={IEEE}
}

@inproceedings{xiao2024neca,
  title={NECA: Neural customizable human avatar},
  author={Xiao, Junjin and Zhang, Qing and Xu, Zhan and Zheng, Wei-Shi},
  booktitle={Proceedings of the IEEE/CVF conference on computer vision and pattern recognition},
  pages={20091--20101},
  year={2024}
}

@inproceedings{wang2024intrinsic,
 title   = {IntrinsicAvatar: Physically Based Inverse Rendering of Dynamic Humans from Monocular Videos via Explicit Ray Tracing},
 author  = {Shaofei Wang and Bo\v{z}idar Anti\'{c} and Andreas Geiger and Siyu Tang},
 booktitle = {Proceedings IEEE Conf. on Computer Vision and Pattern Recognition (CVPR)},
 year    = {2024}
}

@article{kerbl20233d,
  title={3D Gaussian Splatting for Real-Time Radiance Field Rendering},
  author={Kerbl, Bernhard and Kopanas, Georgios and Leimkuehler, Thomas and Drettakis, George},
  journal={ACM Transactions on Graphics (TOG)},
  volume={42},
  number={4},
  pages={1--14},
  year={2023},
  publisher={ACM New York, NY, USA}
}

@article{zhao2025surfel,
  title={Surfel-based Gaussian inverse rendering for fast and relightable dynamic human reconstruction from monocular videos},
  author={Zhao, Yiqun and Wu, Chenming and Huang, Binbin and Zhi, Yihao and Zhao, Chen and Wang, Jingdong and Gao, Shenghua},
  journal={IEEE Transactions on Pattern Analysis and Machine Intelligence},
  year={2025},
  publisher={IEEE}
}

@inproceedings{jiang2025dnf,
  title={DNF-Avatar: Distilling neural fields for real-time animatable avatar relighting},
  author={Jiang, Zeren and Wang, Shaofei and Tang, Siyu},
  booktitle={Proceedings of the IEEE/CVF International Conference on Computer Vision},
  pages={6383--6394},
  year={2025}
}

@article{li2023animatable,
  title={Animatable and relightable gaussians for high-fidelity human avatar modeling},
  author={Li, Zhe and Sun, Yipengjing and Zheng, Zerong and Wang, Lizhen and Zhang, Shengping and Liu, Yebin},
  journal={arXiv preprint arXiv:2311.16096},
  year={2023}
}

@inproceedings{wang2025relightable,
  title={Relightable full-body gaussian codec avatars},
  author={Wang, Shaofei and Simon, Tomas and Santesteban, Igor and Bagautdinov, Timur and Li, Junxuan and Agrawal, Vasu and Prada, Fabian and Yu, Shoou-I and Nalbone, Pace and Gramlich, Matt and others},
  booktitle={Proceedings of the Special Interest Group on Computer Graphics and Interactive Techniques Conference Conference Papers},
  pages={1--12},
  year={2025}
}

@article{zhan2025interactive,
  title={Interactive rendering of relightable and animatable gaussian avatars},
  author={Zhan, Youyi and Shao, Tianjia and Wang, He and Yang, Yin and Zhou, Kun},
  journal={IEEE Transactions on Visualization and Computer Graphics},
  year={2025},
  publisher={IEEE}
}

@inproceedings{choi2025relightable,
  title={Relightable and Dynamic Gaussian Avatar Reconstruction from Monocular Video},
  author={Choi, Seonghwa and Choi, Moonkyeong and Jang, Mingyu and Kim, Jaekyung and Cai, Jianfei and Cheng, Wen-Huang and Lee, Sanghoon},
  booktitle={Proceedings of the 33rd ACM International Conference on Multimedia},
  pages={7405--7414},
  year={2025}
}

@inproceedings{dai2024high,
  title={High-quality surface reconstruction using gaussian surfels},
  author={Dai, Pinxuan and Xu, Jiamin and Xie, Wenxiang and Liu, Xinguo and Wang, Huamin and Xu, Weiwei},
  booktitle={ACM SIGGRAPH 2024 conference papers},
  pages={1--11},
  year={2024}
}

@article{loper2015smpl,
  title={SMPL: a skinned multi-person linear model},
  author={Loper, Matthew and Mahmood, Naureen and Romero, Javier and Pons-Moll, Gerard and Black, Michael J},
  journal={ACM Transactions on Graphics (TOG)},
  volume={34},
  number={6},
  pages={1--16},
  year={2015},
  publisher={ACM New York, NY, USA}
}

@inproceedings{pavlakos2019smplx,
  title = {Expressive Body Capture: {3D} Hands, Face, and Body from a Single Image},
  author = {Pavlakos, Georgios and Choutas, Vasileios and Ghorbani, Nima and Bolkart, Timo and Osman, Ahmed A. A. and Tzionas, Dimitrios and Black, Michael J.},
  booktitle = {Proceedings IEEE Conf. on Computer Vision and Pattern Recognition (CVPR)},
  pages     = {10975--10985},
  year = {2019}
}

@article{jiang2025uv,
  title={Uv gaussians: Joint learning of mesh deformation and gaussian textures for human avatar modeling},
  author={Jiang, Yujiao and Liao, Qingmin and Li, Xiaoyu and Ma, Li and Zhang, Qi and Zhang, Chaopeng and Lu, Zongqing and Shan, Ying},
  journal={Knowledge-Based Systems},
  volume={320},
  pages={113470},
  year={2025},
  publisher={Elsevier}
}

@inproceedings{burov2021dynamic,
  title={Dynamic surface function networks for clothed human bodies},
  author={Burov, Andrei and Nie{\ss}ner, Matthias and Thies, Justus},
  booktitle={Proceedings of the IEEE/CVF International Conference on Computer Vision},
  pages={10754--10764},
  year={2021}
}

@inproceedings{alldieck2018video,
  title={Video based reconstruction of 3d people models},
  author={Alldieck, Thiemo and Magnor, Marcus and Xu, Weipeng and Theobalt, Christian and Pons-Moll, Gerard},
  booktitle={Proceedings of the IEEE Conference on Computer Vision and Pattern Recognition},
  pages={8387--8397},
  year={2018}
}

@inproceedings{alldieck2018detailed,
  title={Detailed human avatars from monocular video},
  author={Alldieck, Thiemo and Magnor, Marcus and Xu, Weipeng and Theobalt, Christian and Pons-Moll, Gerard},
  booktitle={2018 International Conference on 3D Vision (3DV)},
  pages={98--109},
  year={2018},
  organization={IEEE}
}

@article{kwon2023deliffas,
  title={Deliffas: Deformable light fields for fast avatar synthesis},
  author={Kwon, Youngjoong and Liu, Lingjie and Fuchs, Henry and Habermann, Marc and Theobalt, Christian},
  journal={Advances in neural information processing systems},
  volume={36},
  pages={40944--40962},
  year={2023}
}

@article{habermann2023hdhumans,
  title={Hdhumans: A hybrid approach for high-fidelity digital humans},
  author={Habermann, Marc and Liu, Lingjie and Xu, Weipeng and Pons-Moll, Gerard and Zollhoefer, Michael and Theobalt, Christian},
  journal={Proceedings of the ACM on Computer Graphics and Interactive Techniques},
  volume={6},
  number={3},
  pages={1--23},
  year={2023},
  publisher={ACM New York, NY, USA}
}

@article{xiang2022dressing,
  title={Dressing avatars: Deep photorealistic appearance for physically simulated clothing},
  author={Xiang, Donglai and Bagautdinov, Timur and Stuyck, Tuur and Prada, Fabian and Romero, Javier and Xu, Weipeng and Saito, Shunsuke and Guo, Jingfan and Smith, Breannan and Shiratori, Takaaki and others},
  journal={ACM Transactions on Graphics (TOG)},
  volume={41},
  number={6},
  pages={1--15},
  year={2022},
  publisher={ACM New York, NY, USA}
}

@article{xu2011video,
  title={Video-based characters: creating new human performances from a multi-view video database},
  author={Xu, Feng and Liu, Yebin and Stoll, Carsten and Tompkin, James and Bharaj, Gaurav and Dai, Qionghai and Seidel, Hans-Peter and Kautz, Jan and Theobalt, Christian},
  journal={TOG},
  volume={30},
  number={4},
  pages={1--10},
  year={2011},
  publisher={ACM New York, NY, USA}
}

@article{zheng2023avatarrex,
  title={Avatarrex: Real-time expressive full-body avatars},
  author={Zheng, Zerong and Zhao, Xiaochen and Zhang, Hongwen and Liu, Boning and Liu, Yebin},
  journal={ACM Transactions on Graphics (TOG)},
  volume={42},
  number={4},
  pages={1--19},
  year={2023},
  publisher={ACM New York, NY, USA}
}

@inproceedings{li2023posevocab,
  title={Posevocab: Learning joint-structured pose embeddings for human avatar modeling},
  author={Li, Zhe and Zheng, Zerong and Liu, Yuxiao and Zhou, Boyao and Liu, Yebin},
  booktitle={ACM SIGGRAPH 2023 conference proceedings},
  pages={1--11},
  year={2023}
}

@inproceedings{wang2022arah,
  title={Arah: Animatable volume rendering of articulated human sdfs},
  author={Wang, Shaofei and Schwarz, Katja and Geiger, Andreas and Tang, Siyu},
  booktitle={European conference on computer vision},
  pages={1--19},
  year={2022},
  organization={Springer}
}

@inproceedings{zheng2022structured,
  title={Structured local radiance fields for human avatar modeling},
  author={Zheng, Zerong and Huang, Han and Yu, Tao and Zhang, Hongwen and Guo, Yandong and Liu, Yebin},
  booktitle={Proceedings of the IEEE/CVF Conference on Computer Vision and Pattern Recognition},
  pages={15893--15903},
  year={2022}
}

@inproceedings{chen2023uv,
  title={Uv volumes for real-time rendering of editable free-view human performance},
  author={Chen, Yue and Wang, Xuan and Chen, Xingyu and Zhang, Qi and Li, Xiaoyu and Guo, Yu and Wang, Jue and Wang, Fei},
  booktitle={Proceedings of the IEEE/CVF Conference on Computer Vision and Pattern Recognition},
  pages={16621--16631},
  year={2023}
}

@article{liu2021neural,
  title={Neural actor: Neural free-view synthesis of human actors with pose control},
  author={Liu, Lingjie and Habermann, Marc and Rudnev, Viktor and Sarkar, Kripasindhu and Gu, Jiatao and Theobalt, Christian},
  journal={ACM transactions on graphics (TOG)},
  volume={40},
  number={6},
  pages={1--16},
  year={2021},
  publisher={ACM New York, NY, USA}
}

@inproceedings{li2024animatable,
  title={Animatable gaussians: Learning pose-dependent gaussian maps for high-fidelity human avatar modeling},
  author={Li, Zhe and Zheng, Zerong and Wang, Lizhen and Liu, Yebin},
  booktitle={Proceedings of the IEEE/CVF conference on computer vision and pattern recognition},
  pages={19711--19722},
  year={2024}
}

@inproceedings{zhan2026high,
  title={High-Fidelity Mobile Avatars with Pruned Local Blendshapes},
  author={Zhan, Youyi and Wang, He and Shao, Tianjia and Zhou, Kun},
  booktitle={Proceedings of the IEEE/CVF Conference on Computer Vision and Pattern Recognition},
  pages={32345--32356},
  year={2026}
}

@article{li2025anigaussian,
  title={AniGaussian: Animatable Gaussian Avatar with Pose-guided Deformation},
  author={Li, Mengtian and Yao, Shengxiang and Kai, Chen and Xie, Zhifeng and Chen, Keyu and Jiang, Yu-Gang},
  journal={arXiv preprint arXiv:2502.19441},
  year={2025}
}

@inproceedings{hu2024gaussianavatar,
  title={Gaussianavatar: Towards realistic human avatar modeling from a single video via animatable 3d gaussians},
  author={Hu, Liangxiao and Zhang, Hongwen and Zhang, Yuxiang and Zhou, Boyao and Liu, Boning and Zhang, Shengping and Nie, Liqiang},
  booktitle={Proceedings of the IEEE/CVF conference on computer vision and pattern recognition},
  pages={634--644},
  year={2024}
}

@inproceedings{pang2024ash,
  title={Ash: Animatable gaussian splats for efficient and photoreal human rendering},
  author={Pang, Haokai and Zhu, Heming and Kortylewski, Adam and Theobalt, Christian and Habermann, Marc},
  booktitle={Proceedings of the IEEE/CVF Conference on Computer Vision and Pattern Recognition},
  pages={1165--1175},
  year={2024}
}

@inproceedings{zielonka2025drivable,
  title={Drivable 3d gaussian avatars},
  author={Zielonka, Wojciech and Bagautdinov, Timur and Saito, Shunsuke and Zollh{\"o}fer, Michael and Thies, Justus and Romero, Javier},
  booktitle={2025 International Conference on 3D Vision (3DV)},
  pages={979--990},
  year={2025},
  organization={IEEE}
}

@inproceedings{qian20243dgs,
  title={3dgs-avatar: Animatable avatars via deformable 3d gaussian splatting},
  author={Qian, Zhiyin and Wang, Shaofei and Mihajlovic, Marko and Geiger, Andreas and Tang, Siyu},
  booktitle={Proceedings of the IEEE/CVF conference on computer vision and pattern recognition},
  pages={5020--5030},
  year={2024}
}

@inproceedings{kocabas2024hugs,
  title={Hugs: Human gaussian splats},
  author={Kocabas, Muhammed and Chang, Jen-Hao Rick and Gabriel, James and Tuzel, Oncel and Ranjan, Anurag},
  booktitle={Proceedings of the IEEE/CVF conference on computer vision and pattern recognition},
  pages={505--515},
  year={2024}
}

@inproceedings{hu2024gauhuman,
  title={Gauhuman: Articulated gaussian splatting from monocular human videos},
  author={Hu, Shoukang and Hu, Tao and Liu, Ziwei},
  booktitle={Proceedings of the IEEE/CVF conference on computer vision and pattern recognition},
  pages={20418--20431},
  year={2024}
}

@inproceedings{zheng2024gps,
  title={Gps-gaussian: Generalizable pixel-wise 3d gaussian splatting for real-time human novel view synthesis},
  author={Zheng, Shunyuan and Zhou, Boyao and Shao, Ruizhi and Liu, Boning and Zhang, Shengping and Nie, Liqiang and Liu, Yebin},
  booktitle={Proceedings of the IEEE/CVF conference on computer vision and pattern recognition},
  pages={19680--19690},
  year={2024}
}

@inproceedings{chen2024meshavatar,
  title={Meshavatar: Learning high-quality triangular human avatars from multi-view videos},
  author={Chen, Yushuo and Zheng, Zerong and Li, Zhe and Xu, Chao and Liu, Yebin},
  booktitle={European Conference on Computer Vision},
  pages={250--269},
  year={2024},
  organization={Springer}
}

@inproceedings{iqbal2023rana,
  title={Rana: Relightable articulated neural avatars},
  author={Iqbal, Umar and Caliskan, Akin and Nagano, Koki and Khamis, Sameh and Molchanov, Pavlo and Kautz, Jan},
  booktitle={Proceedings of the IEEE/CVF International Conference on Computer Vision},
  pages={23142--23153},
  year={2023}
}

@inproceedings{chen2022relighting4d,
  title={Relighting4d: Neural relightable human from videos},
  author={Chen, Zhaoxi and Liu, Ziwei},
  booktitle={European conference on computer vision},
  pages={606--623},
  year={2022},
  organization={Springer}
}

@inproceedings{ji2022geometry,
  title={Geometry-aware single-image full-body human relighting},
  author={Ji, Chaonan and Yu, Tao and Guo, Kaiwen and Liu, Jingxin and Liu, Yebin},
  booktitle={European Conference on Computer Vision},
  pages={388--405},
  year={2022},
  organization={Springer}
}

@article{pandey2021total,
  title={Total relighting: learning to relight portraits for background replacement.},
  author={Pandey, Rohit and Orts-Escolano, Sergio and Legendre, Chloe and Haene, Christian and Bouaziz, Sofien and Rhemann, Christoph and Debevec, Paul E and Fanello, Sean Ryan},
  journal={ACM Trans. Graph.},
  volume={40},
  number={4},
  pages={43--1},
  year={2021}
}

@article{kanamori2018relighting,
  title={Relighting humans: occlusion-aware inverse rendering for full-body human images},
  author={Kanamori, Yoshihiro and Endo, Yuki},
  journal={ACM Transactions on Graphics (TOG)},
  volume={37},
  number={6},
  pages={1--11},
  year={2018},
  publisher={ACM New York, NY, USA}
}

@inproceedings{zhou2019deep,
  title={Deep single-image portrait relighting},
  author={Zhou, Hao and Hadap, Sunil and Sunkavalli, Kalyan and Jacobs, David W},
  booktitle={Proceedings of the IEEE/CVF international conference on computer vision},
  pages={7194--7202},
  year={2019}
}

@article{wang2020single,
  title={Single image portrait relighting via explicit multiple reflectance channel modeling},
  author={Wang, Zhibo and Yu, Xin and Lu, Ming and Wang, Quan and Qian, Chen and Xu, Feng},
  journal={ACM Transactions on Graphics (ToG)},
  volume={39},
  number={6},
  pages={1--13},
  year={2020},
  publisher={ACM New York, NY, USA}
}

@article{sun2019single,
  title={Single image portrait relighting},
  author={Sun, Tiancheng and Barron, Jonathan T and Tsai, Yun-Ta and Xu, Zexiang and Yu, Xueming and Fyffe, Graham and Rhemann, Christoph and Busch, Jay and Debevec, Paul and Ramamoorthi, Ravi},
  journal={ACM Trans. Graph},
  volume={38},
  number={4},
  pages={1--12},
  year={2019}
}

@inproceedings{carbonera2024relightable,
  title={Relightable neural actor with intrinsic decomposition and pose control},
  author={Carbonera Luvizon, Diogo and Golyanik, Vladislav and Kortylewski, Adam and Habermann, Marc and Theobalt, Christian},
  booktitle={European Conference on Computer Vision},
  pages={465--483},
  year={2024},
  organization={Springer}
}

@article{zeng2025rega,
  title={ReGA: Relighting Dynamic Gaussian Avatars from Sparse Views},
  author={Zeng, Lingzhe and Li, Wensheng and Zheng, Rongbin and Gao, Chengying},
  journal={IEEE Transactions on Circuits and Systems for Video Technology},
  year={2025},
  publisher={IEEE}
}

@inproceedings{saito2024relightable,
  title={Relightable gaussian codec avatars},
  author={Saito, Shunsuke and Schwartz, Gabriel and Simon, Tomas and Li, Junxuan and Nam, Giljoo},
  booktitle={Proceedings of the IEEE/CVF conference on computer vision and pattern recognition},
  pages={130--141},
  year={2024}
}

@inproceedings{gao2024relightable,
  title={Relightable 3d gaussians: Realistic point cloud relighting with brdf decomposition and ray tracing},
  author={Gao, Jian and Gu, Chun and Lin, Youtian and Li, Zhihao and Zhu, Hao and Cao, Xun and Zhang, Li and Yao, Yao},
  booktitle={European Conference on Computer Vision},
  pages={73--89},
  year={2024},
  organization={Springer}
}

@inproceedings{liang2024gs,
  title={Gs-ir: 3d gaussian splatting for inverse rendering},
  author={Liang, Zhihao and Zhang, Qi and Feng, Ying and Shan, Ying and Jia, Kui},
  booktitle={Proceedings of the IEEE/CVF Conference on Computer Vision and Pattern Recognition},
  pages={21644--21653},
  year={2024}
}

@article{chen2024gi,
  title={Gi-gs: Global illumination decomposition on gaussian splatting for inverse rendering},
  author={Chen, Hongze and Lin, Zehong and Zhang, Jun},
  journal={arXiv preprint arXiv:2410.02619},
  year={2024}
}

@inproceedings{wang2023styleavatar,
  title={Styleavatar: Real-time photo-realistic portrait avatar from a single video},
  author={Wang, Lizhen and Zhao, Xiaochen and Sun, Jingxiang and Zhang, Yuxiang and Zhang, Hongwen and Yu, Tao and Liu, Yebin},
  booktitle={ACM SIGGRAPH 2023 Conference Proceedings},
  pages={1--10},
  year={2023}
}

@inproceedings{burley2012physically,
  title={Physically-based shading at disney},
  author={Burley, Brent and Studios, Walt Disney Animation},
  booktitle={Acm siggraph},
  volume={2012},
  number={2012},
  pages={1--7},
  year={2012},
  organization={vol. 2012}
}

@article{rendering2015physically,
  title={Physically-based rendering},
  author={Rendering, Why Physically-Based},
  journal={Procedia IUTAM},
  volume={13},
  number={127-137},
  pages={3},
  year={2015},
  publisher={Elsevier}
}

@article{bavoil2008screen,
  title={Screen space ambient occlusion},
  author={Bavoil, Louis and Sainz, Miguel},
  journal={NVIDIA developer information: http://developers. nvidia. com},
  volume={6},
  number={2},
  pages={6},
  year={2008}
}

@article{paszke2019pytorch,
  title={Pytorch: An imperative style, high-performance deep learning library},
  author={Paszke, Adam and Gross, Sam and Massa, Francisco and Lerer, Adam and Bradbury, James and Chanan, Gregory and Killeen, Trevor and Lin, Zeming and Gimelshein, Natalia and Antiga, Luca and others},
  journal={Advances in neural information processing systems},
  volume={32},
  year={2019}
}

@inproceedings{zwicker2001ewa,
  title={Ewa volume splatting},
  author={Zwicker, Matthias and Pfister, Hanspeter and Van Baar, Jeroen and Gross, Markus},
  booktitle={Proceedings Visualization, 2001. VIS'01.},
  pages={29--538},
  year={2001},
  organization={IEEE}
}

@article{wang2004image,
  title={Image quality assessment: from error visibility to structural similarity},
  author={Wang, Zhou and Bovik, Alan C and Sheikh, Hamid R and Simoncelli, Eero P},
  journal={IEEE transactions on image processing},
  volume={13},
  number={4},
  pages={600--612},
  year={2004},
  publisher={IEEE}
}

@inproceedings{zhang2018unreasonable,
  title={The unreasonable effectiveness of deep features as a perceptual metric},
  author={Zhang, Richard and Isola, Phillip and Efros, Alexei A and Shechtman, Eli and Wang, Oliver},
  booktitle={Proceedings of the IEEE conference on computer vision and pattern recognition},
  pages={586--595},
  year={2018}
}

@article{heusel2017gans,
  title={Gans trained by a two time-scale update rule converge to a local nash equilibrium},
  author={Heusel, Martin and Ramsauer, Hubert and Unterthiner, Thomas and Nessler, Bernhard and Hochreiter, Sepp},
  journal={Advances in neural information processing systems},
  volume={30},
  year={2017}
}

@article{icsik2023humanrf,
  title={Humanrf: High-fidelity neural radiance fields for humans in motion},
  author={I{\c{s}}{\i}k, Mustafa and R{\"u}nz, Martin and Georgopoulos, Markos and Khakhulin, Taras and Starck, Jonathan and Agapito, Lourdes and Nie{\ss}ner, Matthias},
  journal={ACM transactions on graphics (TOG)},
  volume={42},
  number={4},
  pages={1--12},
  year={2023},
  publisher={ACM New York, NY, USA}
}

@article{walter2007microfacet,
  title={Microfacet models for refraction through rough surfaces.},
  author={Walter, Bruce and Marschner, Stephen R and Li, Hongsong and Torrance, Kenneth E},
  journal={Rendering techniques},
  volume={2007},
  pages={18th},
  year={2007}
}

@inproceedings{huang20242d,
  title={2d gaussian splatting for geometrically accurate radiance fields},
  author={Huang, Binbin and Yu, Zehao and Chen, Anpei and Geiger, Andreas and Gao, Shenghua},
  booktitle={ACM SIGGRAPH 2024 conference papers},
  pages={1--11},
  year={2024}
}

@article{kay1986ray,
  title={Ray tracing complex scenes},
  author={Kay, Timothy L and Kajiya, James T},
  journal={ACM SIGGRAPH computer graphics},
  volume={20},
  number={4},
  pages={269--278},
  year={1986},
  publisher={ACM New York, NY, USA}
}

@article{liu2024deepseek,
  title={Deepseek-v3 technical report},
  author={Liu, Aixin and Feng, Bei and Xue, Bing and Wang, Bingxuan and Wu, Bochao and Lu, Chengda and Zhao, Chenggang and Deng, Chengqi and Zhang, Chenyu and Ruan, Chong and others},
  journal={arXiv preprint arXiv:2412.19437},
  year={2024}
}

@article{achiam2023gpt,
  title={Gpt-4 technical report},
  author={Achiam, Josh and Adler, Steven and Agarwal, Sandhini and Ahmad, Lama and Akkaya, Ilge and Aleman, Florencia Leoni and Almeida, Diogo and Altenschmidt, Janko and Altman, Sam and Anadkat, Shyamal and others},
  journal={arXiv preprint arXiv:2303.08774},
  year={2023}
}
